\documentclass[english,10pt,letterpaper]{article}
\usepackage[T1]{fontenc}
\usepackage{graphicx}
\usepackage{mathtools}
\usepackage{amssymb}
\usepackage{amsthm}
\usepackage{xcolor}
\usepackage{babel}
\usepackage[hidelinks]{hyperref}
\usepackage{dsfont}
\usepackage{subfig}
\usepackage[left=2.4cm,right=2.4cm,top=2cm,bottom=2cm]{geometry}
\usepackage{orcidlink}

\newcommand{\ket}[1]{\left| #1 \right\rangle}
\newcommand{\bra}[1]{\left\langle #1 \right|}
\newcommand{\bk}[2]{\left\langle #1 | #2 \right\rangle}
\newcommand{\bkf}[3]{\left\langle #1 \right| #2 \left| #3 \right\rangle}
\newcommand{\pj}[2]{ \left| #1 \right\rangle \left\langle #2 \right|}

\begin{document}
	
	\title{$N$-bein formalism for the parameter space of quantum geometry}
	
		\author{Jorge Romero \orcidlink{0000-0001-8258-6647}, Carlos A. Velasquez \orcidlink{0009-0003-1513-5098}, and J David Vergara \orcidlink{0000-0002-8615-761X} \footnote{Corresponding author} }
	\date{}
	
	\maketitle	
	
	\vspace{-1cm}
	\begin{center}
		\small
		Departamento de F\'{i}sica de Altas Energ\'{i}as, Instituto de Ciencias Nucleares, Universidad Nacional Aut\'{o}noma de M\'{e}xico, Apartado Postal 70-543, Ciudad de M\'{e}xico, 04510, Mexico.\\
	\end{center}
	
	\begin{flushleft}
		Email: jorge.romero@correo.nucleares.unam.mx, carlos.velasquez@correo.nucleares.unam.mx, and vergara@nucleares.unam.mx\\[0.5cm]
	\end{flushleft}
	
	\begin{abstract}
		This work introduces a geometrical object that generalizes the quantum geometric tensor; we call it $N$-bein. Analogous to the vielbein (orthonormal frame) used in the Cartan formalism, the $N$-bein behaves like a ``square root'' of the quantum geometric tensor. Using it, we present a quantum geometric tensor of two states that measures the possibility of moving from one state to another after two consecutive parameter variations. This new tensor determines the commutativity of such variations through its anti-symmetric part. In addition, we define a connection different from the Berry connection, and combining it with the $N$-bein allows us to introduce a notion of torsion and curvature \`{a} la Cartan that satisfies the Bianchi identities. Moreover, the torsion coincides with the anti-symmetric part of the two-state quantum geometric tensor previously mentioned, and thus, it is related to the commutativity of the parameter variations. We also describe our formalism using differential forms and discuss the possible physical interpretations of the new geometrical objects. Furthermore, we define different gauge invariants constructed from the geometrical quantities introduced in this work, resulting in new physical observables. Finally, we present two examples to illustrate these concepts: a harmonic oscillator and a generalized oscillator, both immersed in an electric field. We found that the new tensors quantify correlations between quantum states that were unavailable by other methods.  
	\end{abstract}
	
	\vspace{2pc}
	\noindent
	August 2024\\
	\noindent
	{\it Keywords}: Quantum geometric tensor, Differential geometry, Berry Connection.

	\section{Introduction}
	
	Introducing geometric ideas into physics has led to a deeper understanding of many physical phenomena. For example, general relativity is a completely geometric description of our spacetime. Furthermore, the usual geometric descriptions in quantum mechanics cannot be directly applied, as the coordinates of phase space are non-commutative. This implies that it is necessary to introduce non-commutative formulations of geometry, such as those in \cite{Connes}. Although mathematically correct, these descriptions obscure the usual geometric character, and it would be desirable to have a description where the usual geometric ideas could be directly applied.
	
	On the other hand, in the context of information theory, there is a whole line of work that allows the inclusion of geometric ideas in the context of theories that acquire their information through statistical methods \cite{Amari}, giving rise to Information Geometry. These ideas have also been applied in the context of quantum mechanics; thus, we can study the geometric structure of the parameter space of the theory. Since this space is commutative, all the usual geometric ideas can be applied directly. In this way, the study of the geometry of the parameter space of quantum systems has increased significantly in recent years \cite{Carollo}, representing a deep study of quantum phase transitions and quantum metrology \cite{Toth, Liu2020}. In this context, the quantum geometric tensor \cite{Provost, Zanardi, Zanardi0708}, the Berry curvature \cite{Berry}, and in the case of mixed states, the Bures metric \cite{Bures, Uhlmann, Jozsa} have been introduced.
	On the other hand, in the context of physics in general, there have been notably several geometric descriptions that have shed light on many physical phenomena, such as the curvature of spacetime caused by matter and energy \cite{Weinberg, Carroll}. All of this is described by Riemannian geometry. However, within mathematics, another notably powerful description of geometry corresponds to Cartan geometry \cite{KN1, KN2, Eguchi, Carroll, Straumann2012general, Hehl7607} .  
	
	Cartan's formulation of geometry generalizes Riemannian geometry in a certain sense. In this case, instead of considering only one essential element, such as the metric in Riemannian geometry, Cartan introduces two primary objects: the vielbein and the affine connection. In this form, one achieves a first-order geometric formalism, as opposed to the second-order nature of Riemannian geometry. With these two elements, two "curvatures" are constructed: the covariant derivative of the connection, which is the Riemann curvature, and the covariant derivative of the vielbein, which is the so-called torsion two-form. In the case of zero torsion, Cartan's geometry reduces to Riemannian geometry, as it is possible to rewrite the connection in terms of derivatives of the vielbein and Christoffel symbols.
	Furthermore, one interesting aspect of Cartan's geometry is that it allows for the coupling of gravity with spinor fields~\cite{Weyl_1950,Kibble1,Kibble2}. Thus, it can be said that the energy-momentum density generates curvature, and the spin density generates torsion. With this motivation in mind, this article aims to introduce Cartan's geometry into the context of quantum information geometry and explore whether these tools provide greater insight into the geometric structure of quantum mechanics. In the following sections, we will use the term $N$-bein instead of vielbein as they refer to the $N$-dimensional parameter space.
	
	The content of the article is as follows: In Sec.~\ref{Sec_e_def}, we introduce the definition of the $N$-bein,  which in the context of quantum chemistry is known as non-adiabatic coupling vector \cite{Tully, Crespo, Kryachko0300},  and in a recent paper it has been considered as a connection \cite{Hetenyi}. Nevertheless, our interpretation as an orthonormal frame coincides with the work of Ref. \cite{Ahn2203}. Thus, we use the $N$-beins to construct the quantum geometric tensor, and, similar to its role in Cartan geometry, it behaves as the "square root" of the aforementioned tensor. In Sec.~\ref{Sec_tsQGT}, we introduce three new concepts in the context of quantum information geometry. The first, which we will call the two-state quantum geometric tensor, will have symmetric and anti-symmetric parts. The symmetric part is equivalent to a metric, as we will show in Sec.~\ref{Sec_df}, that it measures the correlation defined by the simultaneous variation of the parameters of the two states. 
	
	On the other hand, the anti-symmetric part, which we will call torsion in analogy to that introduced in Cartan geometry, measures the difference between the simultaneous projections of two orthogonal states. However, unlike the symmetric and anti-symmetric parts of the quantum geometric tensor, which are real, considering the quantum geometric tensor of two states, we will obtain that both quantities can contain a real and an imaginary part.  Sec.~\ref{Sec_Cartan} shows that the torsion has an equivalent definition closely related to that in Cartan geometry; it is the covariant derivative of the $N$-bein. Sec.~\ref{Sec_df} considers how to rewrite our formalism in terms of differential forms. This will allow us to see clearly what the parts of the quantum metric tensor and the Berry curvature consist of and what new combinations contain the torsion $T^{(n,m)}_{ij}$  and the symmetric part  ${\mathcal G}^{(n,m)}_{ij}$. Based on this, we give a geometric interpretation of both quantities as well as their possible physical meaning. Moreover, in Sec.~\ref{Sec_invs}, we discuss how to construct gauge invariants using the new quantities, and thus, obtaining new physical observables. Sec.~\ref{Sec_ejem} contains two examples of quantum mechanical systems where we calculate all the quantities introduced and analyze what new information they provide about the quantum system. Finally, in Sec.~\ref{Sec_concl}, we give the conclusions of our work.

	\section{$N$-bein definition}
	\label{Sec_e_def}
	
	Let $\mathcal{H}$ be the Hilbert space with the orthonormal basis $\{\ket{n}\}_{n=0}^{\infty}$, where the states $\ket{n} = \ket{n(\lambda)}$ depend on the set of $N$ real adiabatic parameters
	\begin{equation}
		\lambda = \{\lambda^{i}| i=1,\dots,N\}.
	\end{equation}
	Thus, each parameter $\lambda^{i}$ is a slowly varying function of time. Sometimes, we omit writing the dependency $\lambda$ on the states $\ket{n}$. Nevertheless, unless stated otherwise, each state depends on $\lambda$. The parameters define a manifold called the parameter space, in which the distance between the states $\ket{n(\lambda + \delta \lambda)}$ and $\ket{n(\lambda)}$ is given by the line element $dl^{2} = g^{(n)}_{ij} d\lambda^{i} d\lambda^{j}$, where $g^{(n)}_{ij}$ is known as the quantum metric tensor \cite{Provost}.  \\ 
	
	On the other hand, in the Hilbert space, the difference between the states $\ket{n(\lambda + \delta \lambda)}$ and $\ket{n(\lambda)}$ at first order is 
	\begin{equation}
		\label{delta_n}
		\ket{\delta n} := \ket{n(\lambda + \delta \lambda)} - \ket{n(\lambda)} \approx \ket{\partial_{i} n} \delta \lambda^{i},
	\end{equation} 
	with $\ket{\partial_{i} n}:= \frac{\partial}{\partial \lambda^{i}} \ket{n(\lambda)}$. Notice the use of Einstein's sum convention for the indices that label the parameters $\lambda^{i}$; we use this convention throughout this work. The change $\ket{\delta n}$ does not necessarily stays on the state $\ket{n(\lambda)}$, because expanding $\ket{\partial_{i} n}$ into the orthonormal basis yields
	\begin{eqnarray}
		\label{din}
		\ket{\partial_{i} n} & = & \left( \sum_{m=0}^{\infty} \pj{m}{m} \right)  \ket{\partial_{i} n} = \bk{n}{\partial_{i}n} \ket{n} + \sum_{m \neq n} \bk{m}{\partial_{i}n} \ket{m}.
	\end{eqnarray}
	Therefore, in general, the variation $\ket{\delta n}$ has a component aligned with the original state $\ket{n(\lambda)}$ as well as in the subspace spanned by the remaining states $\ket{m(\lambda)} \neq \ket{n(\lambda)}$. In the last equality, we identify the Berry connection $A^{(n)}_{i} := \mathrm{i} \bk{n}{\partial_{i} n}$ as the projection onto the state $\ket{n}$, similarly, we define
	\begin{equation}
		\label{e_def}
		e^{(n)}_{i\; m} := \mathrm{i} \bk{m}{\partial_{i}n}, \qquad\qquad m \neq n,
	\end{equation}
	which we call $N$-bein. We opted for the name $N$-bein similar to the name vielbein (many legs) to account for the fact that $e^{(n)}_{i\; m}$ has $N$ ``legs'', one for each of the parameters $\lambda^{i}$. \\
	
	The $N$-bein is our principal tool to describe the parameter space. With it, we derive the quantum geometric tensor, composed by the metric of the space parameter $g^{(n)}_{ij}$ and the curvature of the Berry connection $F^{(n)}_{ij}$; and we also define new geometrical quantities (tensors) that complement the study of the parameter space. We introduce these new objects to study the effects of the change $\lambda \rightarrow \lambda + \delta \lambda$ in two different states $\ket{n(\lambda)}$ and $\ket{m(\lambda)}$. In turn, some of the new tensors present properties akin to Cartan's geometry formalism. We elaborate more on the new geometrical quantities in the upcoming sections. In the meantime, see \ref{Ap_prop} for some identities involving the $N$-bein and other important tensors used in this work.\\
	
\begin{figure}%
	\centering
	\subfloat[]{\includegraphics[scale=0.9]{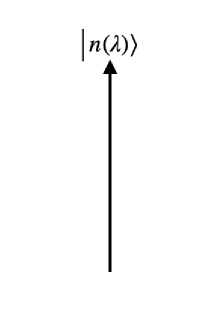}  \label{f_state}}%
	\qquad
	\subfloat[]{\includegraphics[scale=0.9]{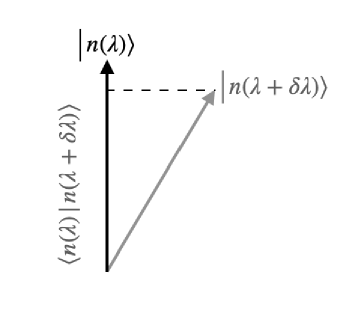}  \label{f_proj1}}%
	\qquad
	\subfloat[]{\includegraphics[scale=0.9]{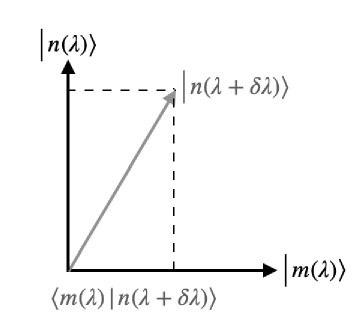}  \label{f_proj2}}%
	\caption{ Projections of a varied state $\ket{n(\lambda + \delta \lambda)}$. (a) Illustration of a state $\ket{n(\lambda)}$. (b) The state after a small variation $\lambda \rightarrow \lambda + \delta \lambda$ and its projection onto the original state $\ket{n(\lambda)}$. (c) The projection of the new state $\ket{n(\lambda + \delta \lambda)}$ onto a different state $\ket{m(\lambda)}.$}
\end{figure}

	Alternatively to its original definition, we can provide a more geometrical interpretation for the $N$-bein. Consider a state $\ket{n(\lambda)}$ represented as an arrow of unit length, Fig.~\ref{f_state}. After a small variation in the parameters $\lambda + \delta \lambda$, part of the new state $\ket{n(\lambda + \delta \lambda)}$ still aligns with the original direction $\ket{n(\lambda)}$, Fig.~\ref{f_proj1}. How much of the varied state $\ket{n(\lambda + \delta \lambda)}$ remains on the original state $\ket{n(\lambda)}$ is measured by the concept of fidelity, $F(\lambda, \lambda + \delta \lambda) = |\bk{n(\lambda)}{n(\lambda + \delta \lambda)}|$. Thus, unless $F(\lambda, \lambda + \delta \lambda)=1$, the new state $\ket{n(\lambda + \delta \lambda)}$ has at least one projection on a state $\ket{m(\lambda)}$ orthogonal to $\ket{n(\lambda)}$, Fig.~\ref{f_proj2}. Such a projection is 
	\begin{equation}
		\label{n_proj}
		\bk{m(\lambda)}{n(\lambda + \delta \lambda)} = \bra{m(\lambda)}\left[ \ket{n(\lambda)} + \ket{\partial_{i} n(\lambda) } \delta \lambda^{i} + \ldots \right] \approx - \mathrm{i} e^{(n)}_{i\; m} \delta \lambda^{i}.
	\end{equation}
	Hence, a non-zero $N$-bein represents the inability of $\ket{n(\lambda)}$ to remain in the same state after a variation of the parameters $\lambda$. Thus, $e^{(n)}_{i\; m}\neq 0$ implies a non-zero probability to go from $\ket{n(\lambda)}$ to $\ket{m(\lambda)}$ or vice versa, because $e^{(n)}_{i\; m}$ is the conjugate of $e^{(m)}_{i\; n}$ [see~\eqref{e_conj}]. Only when $F(\lambda, \lambda + \delta \lambda)=1$, we have $e^{(n)}_{i\; m} = 0$ for all $m$ and no change of state occurs. This reason gives the name ``non-adiabatic coupling vector'' to $e^{(n)}_{i\; m}$ in the context of quantum chemistry~\cite{Kryachko0300}. However, we maintain the name of ``$N$-bein'' because of its geometric relevance close to that of Cartan's geometry. \\

	It is important to remark that for two orthonormal states in the Hilbert space, $\ket{n(\lambda)}$ and $\ket{m(\lambda)}$, the variation $\lambda \rightarrow \lambda + \delta \lambda$ does not alter the orthonormality between the new states $\ket{n(\lambda + \delta \lambda)}$ and $\ket{m(\lambda + \delta \lambda)}$. They will remain orthogonal to each other with norm equal to one (as long as the variation is small enough). The non-zero probability to change state is only with respect to the original unperturbed basis.\\
	
	In the study of quantum geometry, one of the most relevant geometrical quantities is the quantum geometric tensor $Q^{(n)}_{ij}$; since it is composed of the fundamental structures that define the parameter space, the metric $g^{(n)}_{ij}$ and the curvature of the Berry connection ($F^{(n)}_{ij}:= \partial_{i} A^{(n)}_{j} - \partial_{j} A^{(n)}_{i}$). Its definition is 
	\begin{equation}
		\label{qmt_def}
		Q^{(n)}_{ij} := \bra{\partial_{i} n} \left(\hat{\mathds{1}} - \pj{n}{n}\right) \ket{\partial_{j} n} = \bkf{\partial_{i} n}{\hat{P}^{(n)}}{\partial_{j} n},
	\end{equation}
	where we also introduced the projector to the subspace orthogonal to $\ket{n}$, $\hat{P}^{(n)}:= \hat{\mathds{1}} - \pj{n}{n}$. Using the resolution to the identity and~\eqref{e_def}, we write the quantum geometric tensor in terms of the $N$-bein as
	\begin{equation}
		\label{qmt_tetrad}
		Q^{(n)}_{ij} = \sum_{m \neq n} \bk{\partial_{i} n}{m} \bk{m}{\partial_{j} n} = \sum_{m \neq n} e^{(n)}_{i\; m}{}^{*} e^{(n)}_{j\; m}.
	\end{equation}
	Thus, the $N$-bein behaves like a ``square root'' of the quantum geometric tensor, similar to the role played by the vierbein (four legs) in the Cartan formulation of general relativity and other gravitational theories, where it is considered the ``square root'' of the spacetime metric  \cite{Straumann2012general, Hehl7607}. The symmetric (real) part of $Q^{(n)}_{ij}$ is the metric $g^{(n)}_{ij}$, whereas its anti-symmetric (imaginary) part is the curvature of the Berry connection $F^{(n)}_{ij}$. In terms of the $N$-bein, these geometric objects are
	\begin{eqnarray}
		\label{g_tetrad}
		g^{(n)}_{ij} &=& \mathrm{Re} \{ Q^{(n)}_{ij}\} =\frac{1}{2}\sum_{m \neq n} \left( e^{(n)}_{i\; m}{}^{*} e^{(n)}_{j\; m} + e^{(n)}_{i\; m} e^{(n)}_{j\; m}{}^{*} \right),\\
		\label{F_tetrad}
		F^{(n)}_{ij} & = & -2 \mathrm{Im} \{ Q^{(n)}_{ij}\} = \mathrm{i} \sum_{m \neq n} \left( e^{(n)}_{i\; m}{}^{*} e^{(n)}_{j\; m} - e^{(n)}_{i\; m} e^{(n)}_{j\; m}{}^{*} \right).
	\end{eqnarray}
	
	Until now, we have not imposed any restrictions on the basis $\{\ket{n}\}_{n=0}^{\infty}$. When the basis define a non-degenerate quantum system that solves the stationary Schr\"{o}dinger equation, $\hat{H}(\lambda) \ket{n(\lambda)} = E_{n}(\lambda) \ket{n(\lambda)}$, the definition of the $N$-bein~\eqref{e_def} could be rearranged to
	\begin{equation}
		\label{e_def_alt}
		e^{(n)}_{i\; m} = \mathrm{i} \frac{\bkf{m}{\partial_{i} \hat{H}}{n}}{E_{n}-E_{m}}.
	\end{equation}
	Thus, for some quantum systems, this alternative expression for $e^{(n)}_{i\; m}$ gives an easier way to compute the $N$-bein. Moreover, when we substitute \eqref{e_def_alt} into \eqref{qmt_def}, we obtain the Zanardi-Giorda-Cozzini formula for the quantum geometric tensor \cite{Zanardi, Zanardi0708}
	\begin{equation}
		Q^{(n)}_{ij} = \sum_{m \neq n} \frac{\bkf{n}{\partial_{i} \hat{H}}{m} \bkf{m}{\partial_{j} \hat{H}}{n}}{\left( E_{n} - E_{m} \right)^{2}}.
	\end{equation}
	
	\section{Two-state quantum geometric tensor}
	\label{Sec_tsQGT}
	
	When we make a change in the parameters that characterize a quantum system, the quantum geometric tensor $Q^{(n)}_{ij}$ provides the geometric structures in a given state $\ket{n}$.  Using the projector $\hat{P}^{(n)}:= \hat{\mathds{1}} - \pj{n}{n}$, we rewrite $Q^{(n)}_{ij}$ as
	\begin{equation}
		Q^{(n)}_{ij} = \bkf{\partial_{i} n}{\hat{P}^{(n)}}{\partial_{j} n}=\left( \bra{\partial_{i} n} \hat{P}^{(n)} \right) \left( \hat{P}^{(n)} \ket{\partial_{j} n} \right).
	\end{equation}
	Thus, the geometry encoded in $Q^{(n)}_{ij}$ for the state $\ket{n}$ is derived from the projections onto the subspace orthogonal to $\ket{n}$. Therefore, with this idea in mind, we want to study simultaneously how two different states change after the variation $\lambda + \delta \lambda$, so, for $m\neq n$, we define the tensor 
	\begin{equation}
		\label{M_def}
		M^{(n,m)}_{ij} := \left( \bra{\partial_{i} m} \hat{P}^{(m)} \right) \left( \hat{P}^{(n)} \ket{\partial_{j} n} \right) = \bra{\partial_{i}m} \left( \hat{\mathds{1}} - \pj{n}{n} - \pj{m}{m} \right) \ket{\partial_{j} n}.
	\end{equation}
	We call $M^{(n,m)}_{ij}$ the two-state quantum geometric tensor. From the last equation, we identify the projector $\hat{P}^{(n,m)}:= \hat{\mathds{1}} - \pj{n}{n} - \pj{m}{m} = \hat{P}^{(m,n)} $ onto the subspace orthogonal to both $\ket{n}$ and $\ket{m}$. In terms of the $N$-bein, the two-state quantum geometric tensor is
	\begin{equation}
		\label{M_tetrad}
	M^{(n,m)}_{ij} = \sum_{l\neq n, m} e^{(m)*}_{i\; l} e^{(n)}_{j\; l} = \sum_{l\neq n, m} e^{(l)}_{i\; m} e^{(n)}_{j\; l} =-\sum_{l\neq n, m} \langle m|\partial_i l\rangle \langle l|\partial_j n\rangle .
	\end{equation}
	
	Notice that $M^{(n,m)}_{ij}$ resembles $Q^{(n)}_{ij}$, with the difference being that we are now considering two different $N$-beins with a common state $\ket{l} \neq \ket{n}, \ket{l} \neq \ket{m}$. Hence, similar to the quantum geometric tensor, we split $M^{(n,m)}_{ij}$ into its symmetric and anti-symmetric part, which we define as
	\begin{eqnarray}
		\mathcal{G}^{(n,m)}_{ij} &:=& \mbox{Sym} \left( M^{(n,m)}_{ij} \right) = \frac{1}{2} M^{(n,m)}_{ij} + \frac{1}{2} M^{(n,m)}_{ji} \nonumber \\
		&=& \frac{1}{2}\bkf{\partial_{i} m}{\hat{P}^{(n,m)}}{\partial_{j}n} + \frac{1}{2} \bkf{\partial_{j} m}{\hat{P}^{(n,m)}}{\partial_{i}n} \nonumber \\
		\label{M_sym}
		&=& \frac{1}{2} \sum_{l\neq n, m} \left( e^{(m)*}_{i\; l} e^{(n)}_{j\; l} + e^{(m)*}_{j\; l} e^{(n)}_{i\; l} \right)
	\end{eqnarray}
	and
	\begin{eqnarray}
		T^{(n,m)}_{ij} &:=& 2 \mathrm{i} \mbox{ASym} \left( M^{(n,m)}_{ij} \right) = \mathrm{i} M^{(n,m)}_{ij} - \mathrm{i} M^{(n,m)}_{ji} \nonumber \\
		&=& \mathrm{i} \bkf{\partial_{i} m}{\hat{P}^{(n,m)}}{\partial_{j}n} - \mathrm{i} \bkf{\partial_{j} m}{\hat{P}^{(n,m)}}{\partial_{i}n} \nonumber \\
		\label{M_asym}
		&=& \mathrm{i} \sum_{l\neq n, m} \left( e^{(m)*}_{i\; l} e^{(n)}_{j\; l} - e^{(m)*}_{j\; l} e^{(n)}_{i\; l} \right).
	\end{eqnarray} 
	Therefore, 
	\begin{equation}
		M^{(n,m)}_{ij} = \mathcal{G}^{(n,m)}_{ij} + \frac{1}{2 \mathrm{i}} T^{(n,m)}_{ij}.
	\end{equation} 
	The additional scalar factor in the anti-symmetric part will be clear in the next section. Also, keep in mind that, contrary to the metric and the curvature derived from the quantum geometric tensor, the quantities $T^{(n,m)}_{ij}$ and $\mathcal{G}^{(n,m)}_{ij}$ are not necessarily real, both are complex tensors made with real parameters. We chose to separate $M^{(n,m)}_{ij}$ in its symmetric and anti-symmetric parts rather than in its real and imaginary parts due to the geometrical meaning of the objects $T^{(n,m)}_{ij}$ and $\mathcal{G}^{(n,m)}_{ij}$, see Sec.~\ref{Sec_df}. It is important to remark that the definition of $M^{(n,m)}_{ij}$ is only for $\ket{n} \neq \ket{m}$. Thus, $M^{(n,m)}_{ij}$ is not a generalization of the quantum geometric tensor, but it is instead a complement in the study of the geometry of the parameter space. \\

    From Eq.~\eqref{M_tetrad}, we can see that the two-state geometric tensor has the following interpretation. Starting in the state $\ket{n}$, we vary the parameter $\lambda^{j}$ and calculate the amplitude to reach the state $\ket{l}$. Now, in this state, we vary the parameter $\lambda^{i}$ and compute the amplitude to arrive at the state $\ket{m}$. Hence, summing over all common states $\ket{l}$ between $\ket{n}$ and $\ket{m}$ allows us to consider every available path from $\ket{n}$ to $\ket{m}$. In this way, we can say that the two-state geometric tensor measures the ability to reach $|m\rangle$ from the state $|n\rangle$ after two consecutive variations  $\delta \lambda^{j}$ and $\delta \lambda^{i}$. Therefore, if $M^{(n,m)}_{ij}$ is symmetric, then the order of the variations is not important. However, when the $T^{(n,m)}_{ij}\neq 0$, $M^{(n,m)}_{ij}$ has an anti-symmetric part and the sequence of variations to reach $\ket{m}$ from $\ket{n}$ is relevant. On the other hand, the tensor $\mathcal{G}^{(n,m)}_{ij}$ is symmetric by definition, so it does not see the order in which we perform the variations.\\
	
	Under this interpretation, the quantum geometric tensor in \eqref{qmt_tetrad} indicates the ability to stay in the same state $\ket{n}$ after two consecutive variations of the parameters, i.e., we go from an initial state $\ket{n}$ to another state $\ket{m}$ and then after the second variation we are back at $\ket{n}$. The sum over all $m \neq n$ amounts to all the possibilities to return to the initial state $\ket{n}$, corresponding to a cyclic variation. Furthermore, in that case, when the quantum geometric tensor has an anti-symmetric part, it indicates if the order of the parameter variations is relevant, which corresponds to a non-vanishing Berry curvature. Hence, $T^{(n,m)}_{ij}$ is analogous to the Berry curvature, which also has a geometrical interpretation. We elaborate on the meaning of $T^{(n,m)}_{ij}$ in the next section.\\

	\section{Torsion and curvature \`{a} la Cartan}
	\label{Sec_Cartan}	
	
	Consider the following transformation 
	\begin{equation}
		\label{trans}
		\ket{n^{\prime}} = e^{\mathrm{i} \alpha_{n}} \ket{n},
	\end{equation}
	where $\alpha_{n}$ is a real phase that depends on $\lambda$, $\alpha_{n} = \alpha_{n}(\lambda)$. This transformation leaves the norm of $\ket{n}$ unaltered and does not modify the expectation value of the physical observables; it is a gauge transformation.	Under this type of transformation, the Berry connection $A^{(n)}_{i}$ transforms as $A^{(n)}_{i}{}^{\prime} = A^{(n)}_{i} - \partial_{i} \alpha_{n}$, and thus  the name connection. On the other hand, the metric $g^{(n)}_{ij}$ and Berry curvature $F^{(n)}_{ij}$ are physical observables because they are gauge invariant. Meanwhile, using~\eqref{trans} in the $N$-bein definition~\eqref{e_def} results in
	\begin{equation}\label{trule}
		e^{(n)}_{i\; m}{}^{\prime} = e^{\mathrm{i} (\alpha_{n} - \alpha_{m})} e^{(n)}_{i\; m}.
	\end{equation}
        Notice that the gauge transformation considers two different transformation rules, one for $\ket{n}$ and one for $\ket{m}$, because in general $\alpha_{n} \neq \alpha_{m}$. Therefore, as long as the phase $\alpha_{n}$ is not the same for every state $\ket{n}$, the $N$-bein transforms analogous to $\ket{n}$ with a relative phase $\alpha_{nm} := \alpha_{n}-\alpha_{m}$; otherwise it is invariant under the gauge transformation.  Furthermore, contrary to the transformation law of the Berry connection, the $N$-bein transformation lacks the additional term with the partial derivative.\\
	
	Now, allow us to analyze the derivative of $e^{(n)}_{i\; m}$ to measure the change of the $N$-bein as the parameters vary. The quantity $\partial_{i} e^{(n)}_{j\; m}$ does not transform neither as the state $\ket{n}$ nor as a connection. However, if we consider the derivative
	\begin{equation}
		\label{def_De}
		D_{i} e^{(n)}_{j\; m} := \partial_{i} e^{(n)}_{j\; m} + \mathrm{i} \left(  A^{(n)}_{i} - A^{(m)}_{i} \right) e^{(n)}_{j\; m},
	\end{equation}
	its transformation law is $(D_{i} e^{(n)}_{j\; m})^{\prime} = e^{\mathrm{i} \alpha_{nm} } D_{i} e^{(n)}_{j\; m}$, i.e., it is the same as the $N$-bein with the same relative phase. Therefore, $D_{i} e^{(n)}_{j\; m}$ is a covariant derivative since it does not alter the gauge transformation of $e^{(n)}_{i\; m}$. Consequently, the object
	\begin{equation}
		\label{Gamma_def}
		\Gamma^{(n,m)}_{i} := A^{(n)}_{i} - A^{(m)}_{i}
	\end{equation}
	transforms as an Abelian connection, $\Gamma^{(n,m)}_{i}{}^{\prime} = \Gamma^{(n,m)}_{i} - \partial_{i} \alpha_{nm}$. Notice that $\Gamma^{(n,m)}_{i}$ is real, and although it is the difference between two connections, it is not a tensor as expected. The reason is that the phase of the gauge transformation $\alpha_{n}$ depends on the state $\ket{n}$. The connection $A^{(n)}_{i}$ is a connection in the state $\ket{n}$ but not on $\ket{m}$; similarly $A^{(m)}_{i}$ is a connection only on $\ket{m}$. Thus, $\Gamma^{(n,m)}_{i}$ is a connection with respect both states $\ket{n}$ and $\ket{m}$, and it is useful to measure the change of $e^{(n)}_{i\; m}$ through two different states $\ket{n}$ and $\ket{m}$.\\
	
	In Cartan geometry with an orthonormal field (or vielbein) and a connection, we can construct the torsion and curvature 2-forms through the Cartan structure equations. We let the treatment in terms of differential forms for Sec.~\ref{Sec_df}; meanwhile, we take $e^{(n)}_{i \; m}$ analogous the orthonormal field and define the components of a torsion-like tensor
	\begin{equation}
		\label{T_def}
		T^{(n,m)}_{ij} := D_{i}e^{(n)}_{j\;m} - D_{j} e^{(n)}_{i\;m}  = \partial_{i} e^{(n)}_{j\; m} - \partial_{j} e^{(n)}_{i\; m} + \mathrm{i} \Gamma^{(n,m)}_{i} e^{(n)}_{j\; m} - \mathrm{i} \Gamma^{(n,m)}_{j} e^{(n)}_{i\; m}.
	\end{equation}
	The torsion $T^{(n,m)}_{ij}$ (we drop suffix ``-like'' from now on) is the covariant derivative of the $N$-bein anti-symmetrized, so it is anti-symmetric in the indices $i$ and $j$, similar to the curvature of the Berry connection. However, it is worth stressing that the torsion is not invariant under the gauge transformation~\eqref{trans}, see \eqref{T_transf}. Also, the torsion is a complex 2-form that is not necessarily real or imaginary. Nonetheless, the torsion measures the change of $e^{(n)}_{i\; m}$ as we vary the parameters $\lambda^{i}$.\\
	
	When we vary the parameters, $e^{(n)}_{i\; m}$ is related with the variation of $\ket{n}$ projected onto $\ket{m}$, and $T^{(n,m)}_{ij}$ measures the change of $e^{(n)}_{i\; m}$. Hence, the  $T^{(n,m)}_{ij}$ provides information about both states $\ket{n}$ and $\ket{m}$ as the parameters vary, which sounds similar the tensor two-state quantum geometric tensor defined in the previous section. This idea is not misleading. In fact, using the definition of the $N$-bein in the term with the derivative and then simplifying with \eqref{din} results in
	\begin{eqnarray}
		\label{T_def2}
		T^{(n,m)}_{ij} & = & \mathrm{i} \sum_{l \neq n, m} \left( e^{(m)}_{i\; l}{}^{*} e^{(n)}_{j\; l} - e^{(m)}_{j\; l}{}^{*} e^{(n)}_{i\; l} \right).
	\end{eqnarray}
	This expression is precisely the anti-symmetric part of the tensor $M^{(n,m)}_{ij}$ defined on~\eqref{M_def}, that is why we use the same letter. Therefore, the torsion provides information about the states $\ket{n}$ and $\ket{m}$ as we vary the parameters $\lambda^{i}$. Furthermore, using~\eqref{e_def_alt}, the expression for the torsion becomes 
	\begin{equation}
		\label{T_def3}
		T^{(n,m)}_{ij} = \mathrm{i} \sum_{l\neq n,m} \left[ \frac{\bkf{m}{\partial_{i} \hat{H}}{l} \bkf{l}{\partial_{j} \hat{H}}{n}}{(E_{m}-E_{l}) (E_{n}-E_{l})}  - \frac{\bkf{m}{\partial_{j} \hat{H}}{l} \bkf{l}{\partial_{i} \hat{H}}{n}}{(E_{m}-E_{l}) (E_{n}-E_{l})} \right].
	\end{equation}
	Both of these expressions, \eqref{T_def2} and \eqref{T_def3}, are simpler to compute and are equivalent to \eqref{T_def}, which has more geometrical meaning because it is related with the covariant derivative of $e^{(n)}_{i\; m}$.\\
	
	Finally, to conclude this section, we define the curvature of the connection $\Gamma^{(n,m)}_{i}$. Since it is an Abelian connection, its curvature is
	\begin{equation}
		\label{def_R}
		R^{(n,m)}_{ij} := \partial_{i} \Gamma^{(n,m)}_{j} - \partial_{j} \Gamma^{(n,m)}_{i}.
	\end{equation}
	Furthermore, using \eqref{Gamma_def} yields
	\begin{equation}
		\label{R_def}
		R^{(n,m)}_{ij} = F^{(n)}_{ij} - F^{(m)}_{ij}.
	\end{equation}
	The curvature $R^{(n,m)}_{ij}$ is real and invariant under the gauge transformation~\eqref{trans}, and it is another measure between two different states.\\

	\section{Formulation with differential forms}
	\label{Sec_df}
	
	We have been calling $e^{(n)}_{i\; m}$ the $N$-bein; however, they are rather the components of the $N$-bein 1-form 
	\begin{equation}
		\boldsymbol{e}^{(n)}_{m} := e^{(n)}_{i\,m} d \lambda^{i} = \mathrm{i} \bk{m}{\partial_{i} n} d\lambda^{i} = \mathrm{i} \bk{m}{dn},
	\end{equation}
	where $d$ stands for the exterior derivative. Equivalently, when the states $\ket{n}$ satisfy the stationary Schr\"{o}dinger equation, we could use~\eqref{e_def_alt} to write the $N$-bein 1-form as 
	\begin{equation}
		\boldsymbol{e}^{(n)}_{m} = \mathrm{i} \frac{\bkf{m}{d \hat{H}}{n}}{E_{n}-E_{m}},
	\end{equation}
	for $d\hat{H}=\partial_{i}\hat{H} d\lambda^{i}$. Regardless of its expression, working with differential forms simplifies the notation and clarifies the geometric structure behind the objects defined. Thus, we use this section to formulate the previous formalism with differential forms to gain new geometric insight.\\
	
	Prior to the analysis, allow us to introduce the tensor product $\otimes$. For two tensors $\mathcal{T}$ and $\mathcal{T}^{\prime}$ of the type ${r}\choose{s}$ and ${r^{\prime}}\choose{s^{\prime}}$, respectively, the product $\mathcal{T} \otimes \mathcal{T}^{\prime}$ is a ${r+r^{\prime}}\choose{s+s^{\prime}}$-type tensor. Moreover, given two 1-forms $\boldsymbol{\alpha}$ and $\boldsymbol{\beta}$, we use the convention $\mbox{Sym}(\boldsymbol{\alpha} \otimes \boldsymbol{\beta}) = (1/2) ( \boldsymbol{\alpha} \otimes \boldsymbol{\beta} + \boldsymbol{\beta} \otimes \boldsymbol{\alpha} )$ and $\mbox{ASym}(\boldsymbol{\alpha} \otimes \boldsymbol{\beta}) = (1/2) ( \boldsymbol{\alpha} \otimes \boldsymbol{\beta} - \boldsymbol{\beta} \otimes \boldsymbol{\alpha} ) =: \boldsymbol{\alpha} \wedge \boldsymbol{\beta}$. The symbol ``$\wedge$'' represents the anti-symmetric tensor product, known as the wedge or exterior product; it maps a $p$-form and a $q$-form into a $(p+q)$-form. \\
	
	Using differential forms, the Berry connection is $\boldsymbol{A}^{(n)}=A^{(n)}_{i} d\lambda^{i}$. Similarly, the quantum geometric tensor is
	\begin{equation}
		\boldsymbol{Q}^{(n)} = Q^{(n)}_{ij} d\lambda^{i} \otimes d\lambda^{j} = \sum_{m \neq n} \boldsymbol{e}^{(n)\ast}_{m} \otimes \boldsymbol{e}^{(n)}_{m},
	\end{equation}
	which is composed by the metric $\boldsymbol{g}^{(n)}=g^{(n)}_{ij} d\lambda^{i} \otimes d\lambda^{j}$ and the curvature of the Berry connection $\boldsymbol{F}^{(n)}= (1/2) F_{ij} d \lambda^{i} \wedge d \lambda^{j} = d \boldsymbol{A}^{(n)}$, namely
	\begin{equation}
		\boldsymbol{Q}^{(n)} = \boldsymbol{g}^{(n)} - \mathrm{i}  \boldsymbol{F}^{(n)}.
	\end{equation}
	Consequently, in terms of the $N$-bein, we obtain
	\begin{eqnarray}
		\boldsymbol{g}^{(n)} &=& \mbox{Sym}\left( \sum_{m \neq n} \boldsymbol{e}^{(n)\ast}_{m} \otimes \boldsymbol{e}^{(n)}_{m} \right) , \\
		\boldsymbol{F}^{(n)} &=&  \mathrm{i} \, \mbox{ASym}\left( \sum_{m \neq n} \boldsymbol{e}^{(n)\ast}_{m} \otimes \boldsymbol{e}^{(n)}_{m} \right) = \mathrm{i} \sum_{m \neq n} \boldsymbol{e}^{(n)\ast}_{m} \wedge \boldsymbol{e}^{(n)}_{m}.
	\end{eqnarray}
	
	To simplify the former expressions, let us define $\boldsymbol{\theta}^{(n)}_{m}$ and $\boldsymbol{\eta}^{(n)}_{m}$ as the real and imaginary parts of $\boldsymbol{e}^{(n)}_{m}$, i.e., we write 
	\begin{equation}
		\boldsymbol{e}^{(n)}_{m} = \boldsymbol{\theta}^{(n)}_{m} + \mathrm{i} \boldsymbol{\eta}^{(n)}_{m},
	\end{equation}	
	where
	\begin{eqnarray}
		\boldsymbol{\theta}^{(n)}_{m} &=& \mathrm{Re}\{\boldsymbol{e}^{(n)}_{m}\} =\mathrm{i}\frac{\bk{m}{dn} - \bk{dn}{m}}{2} =\mathrm{i} \frac{\bkf{m}{d \hat{H}}{n} - \bkf{n}{d \hat{H}}{m}}{2 (E_{n}-E_{m})}, \\
		\boldsymbol{\eta}^{(n)}_{m} &=& \mathrm{Im}\{\boldsymbol{e}^{(n)}_{m}\} = \frac{ \bk{m}{dn} + \bk{dn}{m}}{2} = \frac{\bkf{m}{d \hat{H}}{n} + \bkf{n}{d \hat{H}}{m}}{2(E_{n}-E_{m})}.
	\end{eqnarray}
	Thus, with $\boldsymbol{\theta}^{(n)}_{m}$ and $\boldsymbol{\eta}^{(n)}_{m}$, the metric $\boldsymbol{g}^{(n)}$ and the curvature $\boldsymbol{F}^{(n)}$ are
	\begin{eqnarray}
		\label{g_forms}
		\boldsymbol{g}^{(n)} &=& \sum_{m \neq n} \left( \boldsymbol{\theta}^{(n)}_{m} \otimes \boldsymbol{\theta}^{(n)}_{m} + \boldsymbol{\eta}^{(n)}_{m} \otimes \boldsymbol{\eta}^{(n)}_{m} \right), \\
		\label{F_forms}
		\boldsymbol{F}^{(n)} &=& -2 \sum_{m \neq n} \boldsymbol{\theta}^{(n)}_{m} \wedge \boldsymbol{\eta}^{(n)}_{m}.
	\end{eqnarray}
	Notice how the metric tensor looks like the square modulus of the $N$-bein, whereas the curvature $\boldsymbol{F}^{(n)}$ is the wedge product between the real and imaginary parts. Hence, if a quantum system only has real or imaginary $N$-beins, then  $\boldsymbol{F}^{(n)}=0$. Also, since $\boldsymbol{F}^{(n)} = d \boldsymbol{A}^{(n)}$, $\sum_{m \neq n} \boldsymbol{\theta}^{(n)}_{m} \wedge \boldsymbol{\eta}^{(n)}_{m}$ is a closed form and $\sum_{m\neq n}  d\boldsymbol{\theta}^{(n)}_{m} \wedge \boldsymbol{\eta}^{(n)}_{m} = \sum_{m\neq n}  \boldsymbol{\theta}^{(n)}_{m} \wedge d\boldsymbol{\eta}^{(n)}_{m} $. \\
	
	Following the same line of thought, we define the 1-form connection
	\begin{equation}
		\boldsymbol{\Gamma}^{(n,m)} := \Gamma^{(n,m)}_{i} d\lambda^{i} = \left( A^{(n)}_{i} - A^{(m)}_{i} \right)  d\lambda^{i} =  \boldsymbol{A}^{(n)} - \boldsymbol{A}^{(m)}.
	\end{equation}
	with its curvature being $\boldsymbol{R}^{(n,m)}:= d \boldsymbol{\Gamma}^{(n,m)}$.  On the other hand, the torsion 2-form is 
	\begin{equation}
		\label{T_form_def}
		\boldsymbol{T}^{(n,m)} := D \boldsymbol{e}^{(n)}_{m} = d \boldsymbol{e}^{(n)}_{m} + \mathrm{i} \boldsymbol{\Gamma}^{(n,m)} \wedge \boldsymbol{e}^{(n)}_{m}.
	\end{equation}
	Similar to \eqref{T_def2}, we have an equivalent expression for the torsion
	\begin{equation}
		\label{T_form_def2}
		\boldsymbol{T}^{(n,m)} = \mathrm{i} \sum_{l \neq n, m}  \boldsymbol{e}^{(m)}_{l}{}^{*} \wedge \boldsymbol{e}^{(n)}_{l}.
	\end{equation} 
	Using both expressions for the torsion, in terms of $\boldsymbol{\theta}^{(n)}_{m}$ and $\boldsymbol{\eta}^{(n)}_{m}$ we have
	\begin{eqnarray}
		\boldsymbol{T}^{(n,m)} &=& d \boldsymbol{\theta}^{(n)}_{m} -  \boldsymbol{\Gamma}^{(n,m)} \wedge \boldsymbol{\eta}^{(n)}_{m} + \mathrm{i} \left(  d \boldsymbol{\eta}^{(n)}_{m} +   \boldsymbol{\Gamma}^{(n,m)} \wedge \boldsymbol{\theta}^{(n)}_{m} \right) \\
		&=& \sum_{l \neq n,m} \Big[ \boldsymbol{\eta}^{(m)}_{l} \wedge \boldsymbol{\theta}^{(n)}_{l} - \boldsymbol{\theta}^{(m)}_{l} \wedge \boldsymbol{\eta}^{(n)}_{l}  + \mathrm{i} \left( \boldsymbol{\theta}^{(m)}_{l} \wedge \boldsymbol{\theta}^{(n)}_{l} + \boldsymbol{\eta}^{(m)}_{l} \wedge \boldsymbol{\eta}^{(n)}_{l} \right)\Big].
	\end{eqnarray}
	The first two terms of the first equation correspond to the real part of $\boldsymbol{T}^{(n,m)}$, and the latter two are its imaginary part. The real part of the torsion resembles the curvature $\boldsymbol{F}^{(n)}$. Although both could be related, be aware that we are considering the case $n\neq m$. On the other hand, the imaginary part of the torsion appears similar to the metric $\boldsymbol{g}^{(n)}$. However, they are not equivalent because $\rm{Im}\{\boldsymbol{T}^{(n,m)}\}$ involves the anti-symmetric  product. As a final remark of the curvature $\boldsymbol{R}^{(n,m)}$ and the torsion $\boldsymbol{T}^{(n,m)}$, notice that they fulfill the Bianchi identities
	\begin{eqnarray}
		D \boldsymbol{T}^{(n,m)} &=& d \boldsymbol{T}^{(n,m)} + \mathrm{i} \boldsymbol{\Gamma}^{(n,m)} \wedge \boldsymbol{T}^{(n,m)} = \mathrm{i} \boldsymbol{R}^{(n,m)} \wedge \boldsymbol{e}^{(n)}_{m}, \\
		d \boldsymbol{R}^{(n,m)} &=& 0.
	\end{eqnarray}

        \begin{figure}
		\centering
		\includegraphics[scale=0.9]{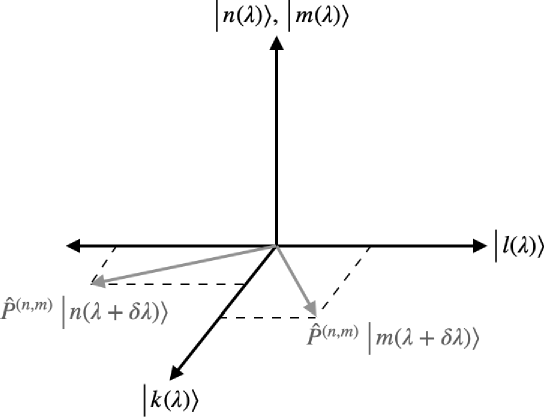}
		\caption{Projections of two varied states, $\ket{n(\lambda + \delta \lambda)}$ and $\ket{m(\lambda + \delta \lambda)}$, onto the subspace orthogonal to $\mbox{span}\{\ket{n(\lambda)}, \ket{m(\lambda)}\}$.}
		\label{f_proj3}
	\end{figure}
 
	To end this section, we explore the tensor $\boldsymbol{M}^{(n,m)}=M^{(n,m)}_{ij} d\lambda^{i} \otimes d\lambda^{j}$, which in terms of the $N$-bein is
	\begin{eqnarray}
		\boldsymbol{M}^{(n,m)} & = &  \sum_{l \neq n, m}  \boldsymbol{e}^{(m)*}_{l} \otimes \boldsymbol{e}^{(n)}_{l} \nonumber \\
		&=& \sum_{l \neq n,m} \Big[ \boldsymbol{\theta}^{(m)}_{l} \otimes \boldsymbol{\theta}^{(n)}_{l} + \boldsymbol{\eta}^{(m)}_{l} \otimes \boldsymbol{\eta}^{(n)}_{l}  + \mathrm{i} \left(  \boldsymbol{\theta}^{(m)}_{l} \otimes \boldsymbol{\eta}^{(n)}_{l} - \boldsymbol{\eta}^{(m)}_{l} \otimes \boldsymbol{\theta}^{(n)}_{l}\right)\Big].
	\end{eqnarray}
	Equivalently, and to gain more geometrical insight, we write $\boldsymbol{M}^{(n,m)}$ as [see \eqref{M_def}] 
	\begin{equation}
		\boldsymbol{M}^{(n,m)} = \left( \bra{dm} \hat{P}^{(m)} \right) \otimes  \left( \hat{P}^{(n)} \ket{dn} \right).
	\end{equation}
        The 1-forms $\ket{dn}$ and $\bra{dm}$ are related with the gradients of $\ket{n}$ and $\bra{m}$, respectively. Thus, using $\hat{P}^{(m)} \hat{P}^{(n)} = \hat{P}^{(n,m)}$ gives the direction of change orthogonal to $\ket{n}$ and $\bra{m}$. Therefore, when both states have a direction of change in common, we can connect them through two parameter variations. If such a connection is not possible, then $\boldsymbol{M}^{(n,m)}$ vanishes. This interpretation coincides with the argument that connecting $\ket{n}$ and $\ket{m}$ through two independent variations requires the existence of an intermediary state $\ket{l}$. Furthermore, recall that $\boldsymbol{M}^{(n,m)} = \boldsymbol{\mathcal{G}}^{(n,m)} - \mathrm{i} \boldsymbol{T}^{(n,m)}$, so we have
	\begin{eqnarray}
		\boldsymbol{\mathcal{G}}^{(n,m)} &=& \mbox{Sym} \left[ \left( \bra{dm} \hat{P}^{(m)} \right) \otimes  \left( \hat{P}^{(n)} \ket{dn} \right) \right], \\
		\boldsymbol{T}^{(n,m)} &=& \mathrm{i} \left( \bra{dm} \hat{P}^{(m)} \right) \wedge \left( \hat{P}^{(n)} \ket{dn} \right).
	\end{eqnarray}
	Therefore, when we study two states $\ket{n(\lambda)}$ and $\ket{m(\lambda)}$ simultaneously, a change in the parameters $\lambda \rightarrow \lambda + \delta \lambda$ does not restrict the new states $\ket{n(\lambda + \delta \lambda)}$ and $\ket{m(\lambda + \delta \lambda )}$ to remain in the subspace spanned by $\ket{n(\lambda)}$ and $\ket{m(\lambda)}$, Fig.~\ref{f_proj3}. It is analogous to the study of only one state (see Sec.~\ref{Sec_e_def}). Thus, $\boldsymbol{\mathcal{G}}^{(n,m)}$ and $\boldsymbol{T}^{(n,m)}$ are measures of the inability of $\ket{n(\lambda + \delta \lambda)}$ and $\ket{m(\lambda + \delta \lambda )}$ to stay in their original subspace. $\boldsymbol{\mathcal{G}}^{(n,m)} \neq 0$ implies that there exist at least one state $\ket{l(\lambda)}$ such that $\boldsymbol{e}^{(n)}_{l} \neq 0$ and $\boldsymbol{e}^{(m)}_{l} \neq 0$. On the other hand, when $\boldsymbol{T}^{(n,m)} \neq 0$ there exists at least two states $\ket{l(\lambda)}$ and $\ket{k(\lambda)}$ for which projections of $\ket{n(\lambda + \delta \lambda)}$ and $\ket{m(\lambda + \delta \lambda )}$ onto such states are different from zero. Therefore, $\boldsymbol{T}^{(n,m)} \neq 0$ is a stronger condition than $\boldsymbol{\mathcal{G}}^{(n,m)} \neq 0$, and it implies that the states $\ket{n(\lambda)}$ and $\ket{m(\lambda)}$ are more likely to leave the subspace they span after a variation in the parameters. Naively, we could say that the torsion is the ``area'' enclosed by 1-forms $\hat{P}^{(m)}\ket{dm}$ and $\hat{P}^{(n)}\ket{dn}$. However, be aware that we are working with complex forms, so the analogy with the area is merely to gain some notion on $\boldsymbol{T}^{(n,m)}$. \\

	\section{Gauge invariants}
	\label{Sec_invs}
	
	In \ref{Ap_prop}, we show some of the properties of the tensors introduced throughout this work. In particular, the transformations laws for $M^{(n,m)}_{ij}$, $\mathcal{G}^{(n,m)}_{ij}$, and $T^{(n,m)}_{ij}$ [given in \eqref{M_transf}, \eqref{G_transf} and \eqref{T_transf}, respectively] imply that they are not gauge invariant unless $\alpha_{nm}=0$. Hence, when $\alpha_{nm}\neq 0$, they are not physical observables. However, we can construct gauge invariants using these new tensors. First, notice that they have the same transformation law as $e^{(n)}_{i \; m}$, see \eqref{trule}. Thus, their real and imaginary parts have similar transformation laws as $\theta^{(n)}_{i \; m}$ and $\eta^{(n)}_{i \; m}$, and they are
	\begin{equation}
        \label{tensor_transf}
		\left( \begin{array}{c} 
			\mathrm{Re} \{ \Xi^{(n,m)}_{ij} \} \\
			\mathrm{Im} \{ \Xi^{(n,m)}_{ij} \}
		\end{array} \right)^{\prime} = \left( \begin{array}{cc} 
			\cos \alpha_{nm} & -\sin \alpha_{nm} \\
			\sin \alpha_{nm} & \cos \alpha_{nm}
		\end{array} \right) \left( \begin{array}{c} 
			\mathrm{Re} \{ \Xi^{(n,m)}_{ij} \} \\
			\mathrm{Im} \{ \Xi^{(n,m)}_{ij}\}
		\end{array} \right),
	\end{equation}
	where $\Xi^{(n,m)}_{ij}$ stands for any of the tensors $M^{(n,m)}_{ij}$, $\mathcal{G}^{(n,m)}_{ij}$, or $T^{(n,m)}_{ij}$.  Therefore, we can form two sets of gauge-invariant tensors. We start with the invariants related to the norm of the vector  $\left( \mathrm{Re} \{ \Xi^{(n,m)}_{ij} \} \quad \mathrm{Im} \{ \Xi^{(n,m)}_{ij} \} \right)^{T}$, they are:
	\begin{eqnarray}
		(N_{M})^{(n,m)}_{ijkl} &:=&  \mathrm{Re} \{ M^{(n,m)}_{ij} \} \mathrm{Re} \{ M^{(n,m)}_{kl} \} + \mathrm{Im} \{ M^{(n,m)}_{ij} \} \mathrm{Im} \{ M^{(n,m)}_{kl} \}, \\
		(N_{\mathcal{G}})^{(n,m)}_{ijkl} &:=& \mathrm{Re} \{ \mathcal{G}^{(n,m)}_{ij}\} \mathrm{Re} \{ \mathcal{G}^{(n,m)}_{kl} \} + \mathrm{Im} \{ \mathcal{G}^{(n,m)}_{ij} \} \mathrm{Im} \{ \mathcal{G}^{(n,m)}_{kl} \}, \\
		(N_{T})^{(n,m)}_{ijkl} &:=& \mathrm{Re} \{ T^{(n,m)}_{ij} \} \mathrm{Re} \{ T^{(n,m)}_{kl} \} + \mathrm{Im} \{ T^{(n,m)}_{ij} \} \mathrm{Im} \{ T^{(n,m)}_{kl} \}.
	\end{eqnarray}
	Then, we have the invariant tensors associated with the area enclosed by $\mathrm{Re} \{ \Xi^{(n,m)}_{ij} \}$ and $\mathrm{Im} \{ \Xi^{(n,m)}_{ij} \}$, namely:
	\begin{eqnarray}
		(A_{M})^{(n,m)}_{ijkl} &:=&  \mathrm{Re} \{ M^{(n,m)}_{ij} \} \mathrm{Im} \{ M^{(n,m)}_{kl} \} - \mathrm{Im} \{ M^{(n,m)}_{ij} \} \mathrm{Re} \{ M^{(n,m)}_{kl} \}, \\
		(A_{\mathcal{G}})^{(n,m)}_{ijkl} &:=& \mathrm{Re} \{ \mathcal{G}^{(n,m)}_{ij}\} \mathrm{Im} \{ \mathcal{G}^{(n,m)}_{kl} \} - \mathrm{Im} \{ \mathcal{G}^{(n,m)}_{ij} \} \mathrm{Re} \{ \mathcal{G}^{(n,m)}_{kl} \}, \\
		(A_{T})^{(n,m)}_{ijkl} &:=& \mathrm{Re} \{ T^{(n,m)}_{ij} \} \mathrm{Im} \{ T^{(n,m)}_{kl} \} - \mathrm{Im} \{ T^{(n,m)}_{ij} \} \mathrm{Re} \{ T^{(n,m)}_{kl} \}.
	\end{eqnarray}
	Observe that the invariant $(N_{T})^{(n,m)}_{ijkl}$ has the same symmetries in the indices as the Riemann curvature tensor. Consequently, when the metric $g^{(n)}_{ij}$ is invertible, we can construct a scalar analogous to the curvature scalar such that it is gauge-invariant. Following this idea, we define the invariants
	\begin{eqnarray}
		\mathcal{N}^{(n,m)}_{M} &:=& 2 g^{(n)ik} g^{(m)jl}(N_{M})^{(n,m)}_{ijkl},\\
		\mathcal{N}^{(n,m)}_{\mathcal{G}} &:=& 2 g^{(n)ik} g^{(m)jl}(N_{\mathcal{G}})^{(n,m)}_{ijkl},\\
		\mathcal{N}^{(n,m)}_{T} &:=& 2 g^{(n)ik} g^{(m)jl}(N_{T})^{(n,m)}_{ijkl},
	\end{eqnarray}
	where $g^{(n)ij}$ is the inverse of $g^{(n)}_{ij}$.\\
	
	Using the $\Xi^{(n,m)}_{ij}$ as in the notation employed before, we summarize the previous results as
	\begin{eqnarray}
		\label{inv_gen1}
		(N_{\Xi})^{(n,m)}_{ijkl} &:=&  \mathrm{Re} \{ \Xi^{(n,m)}_{ij} \} \mathrm{Re} \{ \Xi^{(n,m)}_{kl} \} + \mathrm{Im} \{ \Xi^{(n,m)}_{ij} \} \mathrm{Im} \{ \Xi^{(n,m)}_{kl} \}, \\
		\label{inv_gen2}
		(A_{\Xi})^{(n,m)}_{ijkl} &:=&  \mathrm{Re} \{ \Xi^{(n,m)}_{ij} \} \mathrm{Im} \{ \Xi^{(n,m)}_{kl} \} - \mathrm{Im} \{ \Xi^{(n,m)}_{ij} \} \mathrm{Re} \{ \Xi^{(n,m)}_{kl} \},
	\end{eqnarray}
	and the scalar invariant being
	\begin{equation}
		\mathcal{N}^{(n,m)}_{\Xi} := 2 g^{(n)ik} g^{(m)jl}(N_{\Xi})^{(n,m)}_{ijkl}.
	\end{equation}
	
	This expression similar to the squared norm of a ${0}\choose{2}$-type tensor in a Riemannian manifold~\cite{BookCastillo2011}, the difference being that in our case we are considering two metrics, one for the state $\ket{n}$ and one for the state $\ket{m}$. Additionally, the scalar invariants satisfy $\mathcal{N}^{(n,m)}_{\Xi}=\mathcal{N}^{(m,n)}_{\Xi}$ for all $\Xi$. Thus, given a pair of states $\ket{n}$ and $\ket{m}$, the invariant does not distinguish between moving from $\ket{n}$ to $\ket{m}$ or vice versa. It only measures a relation between two states.\\
	
	To gain some notion on the meaning $\mathcal{N}^{(n,m)}_{\Xi}$, naively we can think of  $\mathcal{N}^{(n,m)}_{\Xi}$ as
	\begin{equation}
		\mathcal{N}^{(n,m)}_{\Xi} \sim \frac{|\boldsymbol{\Xi}^{(n,m)}|^{2}}{\boldsymbol{g}^{(n)} \boldsymbol{g}^{(m)}}
	\end{equation}
	Thus, we are comparing the squared modulus of a complex tensor $\boldsymbol{\Xi}^{(n,m)}$ with two  metrics. The tensor measures the connection between two different states $\ket{n}$ and $\ket{m}$, whereas the metrics are independent among each other. Therefore, when $\mathcal{N}^{(n,m)}_{\Xi}>1$, the effect connecting both states is greater than the independent variations of each state. On the other hand, $\mathcal{N}^{(n,m)}_{\Xi}<1$ implies that the independent variations on each the states are more important than the connection between them. Finally, the case $\mathcal{N}^{(n,m)}_{\Xi}=1$ indicates that both quantities have the same significance and the tensor $\boldsymbol{\Xi}^{(n,m)}$ is ``normalized''.\\
	
	Furthermore, for the sake of completeness, let $\Theta^{(n,m)}_{ij}$ also represent any of tensors $M^{(n,m)}_{ij}$, $\mathcal{G}^{(n,m)}_{ij}$, or $T^{(n,m)}_{ij}$, then we can construct more invariants in the form:
	\begin{eqnarray}
		\label{inv_full1}
		(N_{\Xi \Theta})^{(n,m)}_{ijkl} &:=&  \mathrm{Re} \{ \Xi^{(n,m)}_{ij} \} \mathrm{Re} \{ \Theta^{(n,m)}_{kl} \} + \mathrm{Im} \{ \Xi^{(n,m)}_{ij} \} \mathrm{Im} \{ \Theta^{(n,m)}_{kl} \}, \\
		\label{inv_full2}
		(A_{\Xi \Theta})^{(n,m)}_{ijkl} &:=&  \mathrm{Re} \{ \Xi^{(n,m)}_{ij} \} \mathrm{Im} \{ \Theta^{(n,m)}_{kl} \} - \mathrm{Im} \{ \Xi^{(n,m)}_{ij} \} \mathrm{Re} \{ \Theta^{(n,m)}_{kl} \},
	\end{eqnarray}
	with an equivalent scalar invariant
	\begin{equation}
		\label{sc_inv_full}
		\mathcal{N}^{(n,m)}_{\Xi \Theta} := 2 g^{(n)ik} g^{(m)jl}(N_{\Xi \Theta})^{(n,m)}_{ijkl}.
	\end{equation}
	When $\Theta^{(n,m)}_{kl} = \Xi^{(n,m)}_{kl}$, these invariants reduce to those in~\eqref{inv_gen1} and~\eqref{inv_gen2}. The symmetries in the indices of these invariants depend on the choice of $M^{(n,m)}_{ij}$, $\mathcal{G}^{(n,m)}_{ij}$, or $T^{(n,m)}_{ij}$. Nevertheless,~\eqref{inv_full1}, \eqref{inv_full2}, and~\eqref{sc_inv_full} represent the different invariants constructed from the tensors defined in this paper, which reflect the effect of varying the parameters $\lambda$ on two different states. Notice that $\mathcal{N}^{(n,m)}_{\Xi \Theta}$ maps two ${0}\choose{2}$-type complex tensors to a scalar function in a Riemannian manifold.\\
	
	\section{Illustrative Examples}
	\label{Sec_ejem}
	
	In this section, we study two quantum systems to help us familiarize ourselves with the ideas and concepts introduced in the previous sections. We work out two examples; both are known quantum systems with the addition of a linear term in the Hamiltonian, which models an electric field. These quantum systems allow us to understand the information encoded in the new tensors we defined.\\
	
	\subsection{Harmonic oscillator with a linear term}
	
	The first example under study is the simple harmonic oscillator with a linear term. This system is described by the Hamiltonian
	\begin{eqnarray}
		\label{H_linear}
		\hat{H} &=& \frac{1}{2} \left( \hat{q}^{2} + Z \hat{p}^{2}  \right) + W \hat{q},
	\end{eqnarray}
	where $Z$ is related to the oscillator's frequency and $W$ is the strength of the electric field. The adiabatic parameters are $\{\lambda^{i}\} = \{W,Z\}$, with $i=1,2$. We do not consider a parameter multiplying the first term since the analysis leads to a degenerate quantum metric $g^{(n)}_{ij}$ (for instance, see Ref.~\cite{GGVmetric2019}). Nevertheless, with the proper rescaling of the parameters, a system with three parameters could acquire the form of~\eqref{H_linear}.\\
	
	The solution of the Schr\"{o}dinger equation $\hat{H} \psi_{n} = E_{n} \psi_{n}$ is well known; it is the solution for the simple harmonic oscillator with a translation in the coordinates. Explicitly,
	\begin{equation}
		\psi_{n}(q;\lambda) = \frac{1}{(Z \hbar^{2})^{1/8}}  \chi_{n} \left[  \frac{q+W}{\sqrt{\hbar} Z^{1/4}}   \right],
	\end{equation}
	with 
	\begin{equation}
		\label{chi_def}
		\chi_{n} (\xi) := \left( 2^{n} n! \sqrt{\pi} \right)^{-1/2} e^{-\xi^{2}/2} H_{n} (\xi)
	\end{equation}
	and $H_{n}(\xi)$ is the Hermite polynomial of degree $n$. We took $\xi=(q+W)/(\sqrt{\hbar} Z^{1/4})$ to simplify the notation.  Furthermore, the functions $\chi_{n} (\xi)$ satisfy the properties
	\begin{eqnarray}
		&\displaystyle{\int} d \xi \chi_{n}(\xi) \chi_{m}(\xi) = \delta_{nm}, \\
		&\frac{d \chi_{n}(\xi)}{d \xi} = \sqrt{\frac{n}{2}} \chi_{n-1} (\xi) - \sqrt{\frac{n+1}{2}} \chi_{n+1} (\xi), \\
		&\xi \chi_{n}(\xi) = \sqrt{\frac{n}{2}} \chi_{n-1} (\xi) + \sqrt{\frac{n+1}{2}} \chi_{n+1} (\xi).
	\end{eqnarray}
	On the other hand, the eigenvalues for the Hamiltonian~\eqref{H_linear} are
	\begin{equation}
		E_{n}  = \hbar \sqrt{Z} \left( n + \frac{1}{2} \right) - \frac{W^{2}}{2},
	\end{equation}
	which are the same as those for the simple harmonic oscillator with a constant translation of $-W^{2}/2$. \\
	
	Once we have the solution for the system, the computation of the $N$-bein is straightforward. We either use the definition~\eqref{e_def} or the formula~\eqref{e_def_alt}. Independently of the path taken, the result will be the same. Here, we take the first approach and write~\eqref{e_def} in the coordinate representation
	\begin{equation}
		e^{(n)}_{i\; m} := \mathrm{i} \int^{\infty}_{-\infty} dq \; \psi_{m}^{\ast} (q; \lambda) \partial_{i} \psi_{n} (q; \lambda).
	\end{equation}
	Then, we find that for a given wave function in the $n$-th state, there are only four non-zero $N$-beins, all of them are pure imaginary functions; they are
	\begin{eqnarray}
		(e^{(n)}_{i\; n-1})^{T} &=& \frac{\mathrm{i} }{Z^{1/4}} \sqrt{\frac{n}{2 \hbar}} \left( \begin{array}{c}
			1 \\
			0
		\end{array} \right),  \qquad \qquad
		(e^{(n)}_{i\; n+1})^{T}=  - \frac{\mathrm{i}}{Z^{1/4}} \sqrt{\frac{n+1}{2 \hbar}} \left( \begin{array}{c}
			1 \\
			0
		\end{array}\right), \\
		(e^{(n)}_{i\; n-2})^{T} &=& -\frac{\mathrm{i} }{8 Z} \sqrt{n (n-1)} \left( \begin{array}{c}
			0 \\
			1
		\end{array} \right), \quad
		(e^{(n)}_{i\; n+2})^{T} = \frac{\mathrm{i} }{8 Z} \sqrt{(n+1)(n+2)} \left( \begin{array}{c}
			0 \\
			1
		\end{array} \right).
	\end{eqnarray}
	The $N$-beins $e^{(n)}_{i\; n \pm 2}$ are related with the simple harmonic oscillator because their first component, associated with the parameter $W$, is zero. Consequently,  $e^{(n)}_{i\; n \pm 1}$ arise due the linear term, since they only have the first component. Thus, $e^{(n)}_{i\; n \pm 1}$ are characteristic of a system with a linear term.\\
	
	Regardless of the $N$-bein, notice that they do not depend on the parameter $W$. Hence, the parameter space described by the system whose Hamiltonian is~\eqref{H_linear} has a gauge symmetry. In particular, $\hat{H}$ is invariant under translations in $W$. Moreover, since the $N$-beins do not depend on $W$, neither will the derived quantities (metric, curvature, torsion, etc.) depend on $W$. On the other hand, the $N$-beins are well defined as long as $Z\neq 0$, which is the case when the system is no longer a harmonic oscillator, but rather it is like a particle trapped in a constant potential without kinetic energy.  \\
	
	Recall that the $N$-beins measure a relation between two states. Thus, when the state under study is $n=0$ or $n=1$, we have $e^{(n)}_{i\; n-2}=0$, consequence that there are no states with a negative $n$. The same happens for $e^{(n)}_{i\; n-1}=0$ when $n=0$. Additionally, given that the $N$-beins are pure imaginary functions, $F^{(n)}=0$ [see~\eqref{F_forms}], which is consistent with the fact that $A^{(n)}_{i} = 0$.	\\
	
	Using~\eqref{qmt_def}, it is straightforward to compute the quantum geometric tensor; the result is
	\begin{eqnarray}
		(Q^{(n)}_{ij}) &=& (e^{(n)}_{i\; n-1})^{\dagger} (e^{(n)}_{j\; n-1}) + (e^{(n)}_{i\; n+1})^{\dagger} (e^{(n)}_{j\; n+1}) + (e^{(n)}_{i\; n-2})^{\dagger} (e^{(n)}_{j\; n-2}) + (e^{(n)}_{i\; n+2})^{\dagger} (e^{(n)}_{j\; n+2}) \nonumber \\
		\label{qmt_ex1}
		&=& \frac{n + \frac{1}{2}}{\hbar \sqrt{Z}} \left( \begin{array}{cc}
			1 & 0    \\
			0 & 0 
		\end{array} \right)
		+ \frac{n^{2} + n + 1}{32 Z^{2}}
		\left( \begin{array}{cc}
			0 & 0  \\
			0 & 1  
		\end{array} \right).
	\end{eqnarray}
	The use of $N$-beins allows us to analyze the construction of the quantum geometric tensor. In particular, the first part of~\eqref{qmt_ex1} corresponds to the $N$-beins $e^{(n)}_{i\; n\pm 1}$, associated with the parameter $W$, and the second matrix (and the factor multiplying it) are distinctive of the simple harmonic oscillator which is due the to the $N$-beins $e^{(n)}_{i\; n \pm 2}$. Furthermore, the quantum geometric tensor is real. Therefore, $g^{(n)}_{ij} = Q^{(n)}_{ij}$ and $F^{(n)}_{ij}=0$ as we have already predicted from the structure of the $N$-beins. Similarly, we mention that the curvature $R^{(n,m)}_{ij}$, defined in~\eqref{R_def} is zero because $\Gamma^{(n,m)}_{i}=0$. The $N$-bein contains the same information as the quantum geometric tensor, and it is easier to compute.\\

    Notice that the determinant of the quantum metric tensor,
	\begin{equation}
		 \det (g^{(n)}_{ij}) = \frac{(2n + 1)(n^{2} + n + 1)}{64 \hbar Z^{5/2}},
	\end{equation}
	vanishes only when $Z \rightarrow \infty$ and diverges for $Z=0$. Thus, in the case where $\det (g^{(n)}_{ij}) \neq 0$, we are able to compute the Christoffel symbols as well as the the Riemannian curvature and the scalar curvature of the parameter space. The latter results in
	\begin{equation}
            \label{e1_R}
		\mathcal{R} = - \frac{4}{n^{2} + n + 1}.
	\end{equation}
	Therefore, in the ground state ($n=0$), the space parameter has a negative constant curvature, and as we move away from the ground state, the space parameter becomes flatter. \\
 
	Continuing with the analysis, first we obtain the symmetric part of the two-state quantum geometric tensor, so using~\eqref{M_sym} yields
	\begin{eqnarray}
		(\mathcal{G}^{(n,n-4)}_{ij}) &=& -\frac{1}{64 Z^{2}} \sqrt{n (n-1) (n-2) (n-3)} \left( \begin{array}{cc}
			0 & 0    \\
			0 & 1 
		\end{array} \right), \\
		(\mathcal{G}^{(n,n-3)}_{ij}) &=& \frac{1}{8 Z^{5/4}} \sqrt{\frac{n (n-1) (n-2)}{2 \hbar}} \left( \begin{array}{cc}
			0 & 1    \\
			1 & 0 
		\end{array} \right), \\
		(\mathcal{G}^{(n,n-2)}_{ij}) &=& -\frac{1}{2 \hbar} \sqrt{\frac{n (n-1)}{Z}} \left( \begin{array}{cc}
			1 & 0    \\
			0 & 0 
		\end{array} \right), \\
		(\mathcal{G}^{(n,n-1)}_{ij}) &=& - \frac{n}{8 Z^{5/4}} \sqrt{\frac{n}{2 \hbar}} \left( \begin{array}{cc}
			0 & 1    \\
			1 & 0 
		\end{array} \right) , \\
		(\mathcal{G}^{(n,n+1)}_{ij}) &=& - \frac{ (n+1)}{8 Z^{5/4}} \sqrt{\frac{n+1}{2 \hbar}} \left( \begin{array}{cc}
			0 & 1    \\
			1 & 0 
		\end{array} \right), \\
		(\mathcal{G}^{(n,n+2)}_{ij}) &=& -\frac{1}{2 \hbar} \sqrt{\frac{(n+1) (n+2)}{Z}} \left( \begin{array}{cc}
			1 & 0    \\
			0 & 0 
		\end{array} \right), \\
		(\mathcal{G}^{(n,n+3)}_{ij}) &=& \frac{1}{8 Z^{5/4}} \sqrt{\frac{(n+1) (n+2) (n+3)}{2 \hbar}} \left( \begin{array}{cc}
			0 & 1    \\
			1 & 0 
		\end{array} \right), \\
		(\mathcal{G}^{(n,n+4)}_{ij}) &=& -\frac{1}{64 Z^{2}} \sqrt{(n+1) (n+2) (n+3) (n+4)} \left( \begin{array}{cc}
			0 & 0    \\
			0 & 1 
		\end{array} \right).
	\end{eqnarray}
	Therefore, for a given state $\ket{n}$ at most it is related with the states $\ket{n\pm 1}$, $\ket{n\pm 2}$, $\ket{n\pm 3}$, and  $\ket{n\pm 4}$. Also, similar to the $N$-bein, we have $\mathcal{G}^{(n,n-a)}_{ij} = 0$ when $n<a$ with $a=1,2,3,4$, because negative states do not exist. Moreover, the structure of the matrices that constitute $\mathcal{G}^{(n,n\pm a)}_{ij}$ allows us to determine the path from the state $\ket{n}$ to the state $\ket{n \pm a}$. For example, to reach the states $\ket{n\pm 2}$, we require two variations on the parameter $W$. On the other hand, we need two consecutive variations on the parameter $Z$ to arrive at the states $\ket{n\pm 4}$. Meanwhile, to connect the state $\ket{n}$ with either $\ket{n \pm 1}$ or $\ket{n \pm 3}$, we require two different variations: one on $W$ and one on $Z$. Whether or not the order of the variations is relevant will be determined by the torsion.\\
	
	Now, for the anti-symmetric part of the two-state tensor or the torsion, we use~\eqref{M_asym} and obtain
	\begin{eqnarray}
		(T^{(n,n-1)}_{ij}) &=& -\frac{\mathrm{i}}{4 Z^{5/4}} \sqrt{\frac{n}{2 \hbar}} \left( \begin{array}{cc}
			0 & -1    \\
			1 & 0 
		\end{array} \right), \\
		(T^{(n,n+1)}_{ij}) &=&  \frac{\mathrm{i}}{4 Z^{5/4}} \sqrt{\frac{n+1}{2 \hbar}} \left( \begin{array}{cc}
			0 & -1    \\
			1 & 0 
		\end{array} \right).
	\end{eqnarray}
	Hence, in the connection between the states $\ket{n}$ and $\ket{n \pm 1}$, the order of the variations of $W$ and $Z$ matters. In contrast, to connect the states  $\ket{n}$ and $\ket{n \pm 3}$ we must vary $W$ and $Z$, but in this case the order is not important. Although none of the geometric quantities depend on the parameter $W$, its presence in the Hamiltonian causes a non-zero torsion. Therefore, the geometry of the space parameter changes due to the linear term, even though it does not depend on the parameter $W$. This is an indication that our new variables $\mathcal{G}^{(n,m)}_{ij}$ and $T^{(n,m)}_{ij}$ observe not only the initial value of the parameters but also the possible changes that they may have. \\

    We finalize this example with a discussion on the scalar invariants $\mathcal{N}^{(n,m)}_{\Xi}$, for $\Xi= T, \, \mathcal{G}$, and $M$. For this example, the invariants do not depend on the parameters. Hence, we plot the value of $\mathcal{N}^{(n,n \pm a)}_{\Xi}$ for different states $\ket{n}$ and $a=1,2,3,4$. Also, keep in mind that for $a=2,3,4$, we have $\mathcal{N}^{(n, n \pm a)}_{M} = \mathcal{N}^{(n,n \pm a)}_{\mathcal{G}}$ because in these cases $T^{(n,n \pm a)}_{ij}=0$.\\
    
     \begin{figure}[h]%
    	\centering
    	\subfloat[ ][]{\includegraphics[scale=0.7]{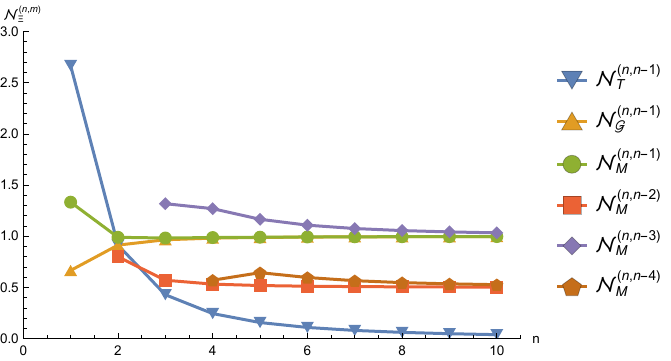}  \label{Nm_e1}}
    	\quad
    	\subfloat[ ][]{\includegraphics[scale=0.7]{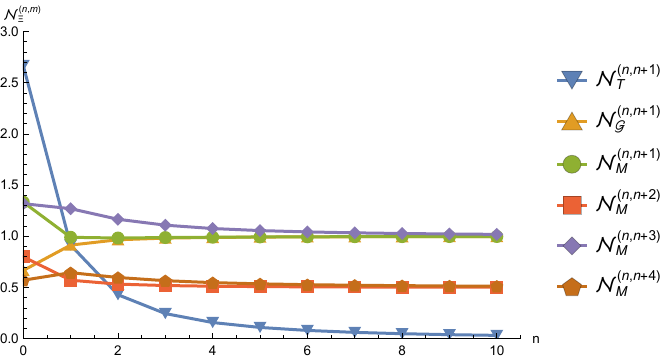}  \label{Np_e1}}
    	\caption{The non-zero invariants $\mathcal{N}^{(n,m)}_{\Xi}$ for $M^{(n,m)}_{ij}$, $T^{(n,m)}_{ij}$, and $\mathcal{G}^{(n,m)}_{ij}$. (a) Graph of the invariants a $\mathcal{N}^{(n,n-a)}_{\Xi}$, with $a=1,2,3,4$ and $\Xi=M,\, T, \, \mathcal{G}$. (b) Graph of the invariants a $\mathcal{N}^{(n,n+a)}_{\Xi}$, with $a=1,2,3,4$ and $\Xi=M,\, T, \, \mathcal{G}$.}
    \end{figure}

	In Fig.~\ref{Nm_e1}, we present the scalar invariants of the form $\mathcal{N}^{(n,n-a)}_{\Xi}$, where the first non-zero point corresponds with the value of $a$ because negative states do not exist and the invariant vanishes. On the other hand, in Fig.~\ref{Np_e1}, we show the scalar invariants of the form $\mathcal{N}^{(n,n+a)}_{\Xi}$. Notice how both graphs provide the same information due to the property $\mathcal{N}^{(n,m)}_{\Xi}=\mathcal{N}^{(m,n)}_{\Xi}$ already discussed. Nonetheless, we present both cases to expose this feature explicitly.\\

	From the plots, we observe that the torsion invariants $\mathcal{N}^{(n,n \pm 1)}_{T}$ approach zero as $n$ grows. This behavior resembles the scalar curvature $\mathcal{R}$ derived in \eqref{e1_R}. Therefore, the relevance of the torsion decreases as the parameter space becomes flatter. It is also interesting to mention that $\mathcal{N}^{(0,1)}_{T}$ has the highest value of all the invariants, and then it declines abruptly. We ignore the cause of such demeanor, but it could be related to the fact that the ground state is a Gaussian state, and moving to the first excited level ($n=1$) results in a quantifiable change. Meanwhile, the rest of the invariants have similar behaviors among them. In particular, we can divide them into two sets. The first set corresponds to the invariants $\mathcal{N}^{(n,n \pm 2)}_{M}$ and $\mathcal{N}^{(n,n \pm 4)}_{M}$, which tend to 1/2 as $n$ increases. On the other hand, in the second set, we have the rest of the invariants, whose values approach the unit when $n$ grows. \\

	\subsection{Generalized harmonic oscillator with a linear term}	
	
	The next example is the generalized harmonic oscillator with a linear term. This quantum system is described by the Hamiltonian
	\begin{eqnarray}
		\label{H_gen_linear}
		\hat{H} &=& \frac{1}{2} \left[ \hat{q}^{2} + Y \left( \hat{q} \hat{p} + \hat{p} \hat{q} \right) + Z \hat{p}^{2}  \right] + W \hat{q},
	\end{eqnarray}
	where $\{\lambda^{i}\}= \{W,Y,Z\}$ are the adiabatic parameters ($i=1,2,3$). The wave function that solves the Schr\"{o}dinger equation involving this Hamiltonian is~\cite{GGVmetric2019}
	\begin{equation}
		\psi_{n}(q;\lambda) = \left( \frac{\omega}{Z \hbar} \right)^{1/4} \chi_{n} \left[ \sqrt{\frac{\omega}{Z \hbar}} \left( q+\frac{WZ}{\omega^{2}}\right) \right] \mbox{Exp} \left( - \frac{\mathrm{i} Y}{2 Z \hbar} q^{2}\right),
	\end{equation}
	being $\omega := \sqrt{Z - Y^{2}}$ the oscillation frequency and $\chi_{n} (\xi)$ is defined back in~\eqref{chi_def}, where in this case we take
	\begin{equation}
		\xi=\sqrt{\frac{\omega}{Z \hbar}} \left( q+\frac{WZ}{\omega^{2}}\right).
	\end{equation}
	On the other hand, the eigenvalues for $\hat{H}$ are
	\begin{equation}
		E_{n} = \hbar \omega \left( n + \frac{1}{2} \right) - \frac{W^{2} Z}{2 \omega^{2}}.
	\end{equation}
	
	Following the same steps as before, we use~\eqref{e_def} and obtain that the only non-zero $N$-beins for a given wave function on the state $n$. They are
	\begin{eqnarray}
		(e^{(n)}_{i\; n-1})^{T} &=&  - W \sqrt{\frac{n }{2 \hbar Z \omega^{5}}}
		\left( \begin{array}{c} 
			0 \\
			Z \\
			-  Y
		\end{array} \right)
		+ \mathrm{i} \sqrt{\frac{n}{2 \hbar Z \omega^{7}}} \left( \begin{array}{c}
			Z \omega^{2}   \\
			2 WYZ\\
			- WY^{2} 
		\end{array} \right),\\ 
		(e^{(n)}_{i\; n+1})^{T} &=&  - W \sqrt{\frac{n+1}{2 \hbar Z \omega^{5}}}
		\left( \begin{array}{c} 
			0 \\
			Z \\
			-  Y
		\end{array} \right)
		- \mathrm{i} \sqrt{\frac{n+1}{2 \hbar Z \omega^{7}}} \left( \begin{array}{c}
			Z \omega^{2}   \\
			2 WYZ\\
			- WY^{2} 
		\end{array} \right),\\
		(e^{(n)}_{i\; n-2})^{T} &=&  \frac{\sqrt{n (n-1)}}{4 Z \omega}
		\left( \begin{array}{c} 
			0 \\
			Z \\
			-Y
		\end{array} \right)
		- \mathrm{i} \frac{\sqrt{n (n-1)}}{8 Z \omega^{2}} \left( \begin{array}{c}
			0  \\
			2 YZ\\
			Z - 2 Y^{2}
		\end{array} \right),\\
		(e^{(n)}_{i\; n+2})^{T} &=&  \frac{\sqrt{(n+1) (n+2)}}{4 Z \omega}
		\left( \begin{array}{c} 
			0 \\
			Z \\
			-Y
		\end{array} \right)
		+ \mathrm{i} \frac{\sqrt{(n+1) (n+2)}}{8 Z \omega^{2}} \left( \begin{array}{c}
			0  \\
			2 YZ\\
			Z - 2 Y^{2}
		\end{array} \right).
	\end{eqnarray}
	As in the former example, we have four different $N$-beins, and all of them are complex 1-forms. Also, the $N$-beins $e^{(n)}_{i\; n\pm 1}$ depend on the parameter $W$, so the metric will no longer be invariant under translations of the parameter $W$. On the other hand, the $N$-beins diverge whether $Z\rightarrow 0$ or $\omega \rightarrow 0$, which translates into the phase-transitions precursors predicted by the quantum geometric tensor.
	
	Furthermore, notice that all the real parts of the $N$-beins are proportional to each other. In fact, they are all proportional to the Berry connection 
	\begin{equation}
		(A^{(n)}_{i})^{T} =  \left( \frac{n+\frac{1}{2}}{2 Z\omega} + \frac{ W^{2}}{2 \hbar \omega^{4}} \right)
		\left( \begin{array}{c} 
			0 \\
			Z \\
			- Y
		\end{array} \right),
	\end{equation}
	and therefore to the connection $\Gamma^{(n,m)}_{i}$
	\begin{equation}
		\label{Gamma_ejem}
		(\Gamma^{(n,m)}_{i})^{T} :=  (A^{(n)}_{i} - A^{(n)}_{i})^{T}  = \frac{n-m}{2 Z\omega} 
		\left( \begin{array}{c} 
			0 \\
			Z \\
			-Y
		\end{array} \right).
	\end{equation}
	Notice that for every energy level $n$, $W$ appears a constant translation in $A^{(n)}_{i}$.  Consequently, $\Gamma^{(n,m)}_{i}$ does not depend on $W$, because it is constructed in a way it eliminates this constant translation.
	
	The quantum geometric tensor could be directly obtained from~\eqref{qmt_def}. However, to simplify its derivation, we use~\eqref{g_forms} and~\eqref{F_forms} to compute $g^{(n)}_{ij}$ and $F^{(n)}_{ij}$, respectively. Thus, we obtain
	\begin{eqnarray}\label{metric}
		(g^{(n)}_{ij}) &=&  \frac{n+\frac{1}{2}}{\hbar \omega^{7}} 
		\left( \begin{array}{ccc}
			Z \omega^{4} & 2 WYZ \omega^{2} & -WY^{2} \omega^{2}  \\
			2 WYZ \omega^{2} & W^{2} Z (3Y^{2} + Z ) &  - W^{2} Y (Y^{2} + Z) \\
			-WY^{2} \omega^{2} & - W^{2} Y (Y^{2} + Z) & W^{2} Y^{2}
		\end{array} \right)  \nonumber \\
		&&
		+ \frac{n^{2} + n + 1}{32 \omega^{4}} \left( \begin{array}{ccc}
			0 & 0 &  0 \\
			0 & 4Z & -2Y \\
			0 & -2Y & 1
		\end{array} \right),
	\end{eqnarray}
	which has a non vanishing determinant unless $Z=0$ and diverges when $\omega\to 0$, given by
	
	\begin{equation}
		\det(g_{ij}^{(n)})=\frac{\left(n+\frac{1}{2} \right)(n^{2}+n+1)Z\left[(n^{2}+n+1) \hbar \omega^{3}+8\left(n+\frac{1}{2} \right)W^{2}Z \right]}{256 \hbar^{2} \omega^{12}},
	\end{equation}
	and for Berry's curvature we have
	
	\begin{eqnarray}
		(F^{(n)}_{ij}) &=& \frac{W}{\hbar \omega^{6}} 
		\left( \begin{array}{ccc}
			0 &  Z \omega^{2} & -Y \omega^{2}  \\
			-Z \omega^{2} & 0 &  - W Y^{2}  \\
			Y \omega^{2} &  W Y^{2} & 0
		\end{array} \right)
		+ \frac{n + \frac{1}{2}}{4 \omega^{3}}
		\left( \begin{array}{ccc}
			0 & 0 &  0 \\
			0 & 0 & -1 \\
			0 & 1 & 0
		\end{array} \right).
	\end{eqnarray}
	The first matrices of $g^{(n)}_{ij}$ and $F^{(n)}_{ij}$ are derived directly from the components of the $N$-beins $e^{(n)}_{i\; n \pm 1}$, which only appear due to the linear term in the Hamiltonian~\eqref{H_gen_linear}. Therefore, it is no coincidence that only these parts of $g^{(n)}_{ij}$ and $F^{(n)}_{ij}$ depend on the parameter $W$. Meanwhile, the curvature of $\Gamma^{(n,m)}_{i}$ is
	\begin{equation}
		(R^{(n,m)}_{ij}) =  F^{(n)}_{ij} - F^{(m)}_{ij} = \frac{n - m}{4 \omega^{3}}
		\left( \begin{array}{ccc}
			0 & 0 &  0 \\
			0 & 0 & -1 \\
			0 & 1 & 0
		\end{array} \right), 
	\end{equation}
	which, just as $\Gamma^{(n,m)}_{i}$, it does not depends on the parameter $W$.\\
	
	
	Continuing with the analysis, we compute the symmetric and anti-symmetric parts of the two-state quantum geometric tensor $M^{(n,m)}_{ij}$. The values of $m$ for which the symmetric tensor $\mathcal{G}^{(n,m)}_{ij}$ is non-zero are
	{\scriptsize	\begin{eqnarray}(\mathcal{G}^{(n,n-4)}_{ij}) &=&  \frac{\sqrt{n (n-1) (n-2) (n-3)}}{64 Z^{2} \omega^{4}} \Bigg[\left(
			\begin{array}{ccc}
				0 & 0 & 0 \\
				0 & 4 Z^2 \left(Z-2 Y^2\right) & 2 Y Z \left(4 Y^2-3 Z\right) \\
				0 & 2 Y Z \left(4 Y^2-3 Z\right) & -8 Y^4+8 Y^2 Z-Z^2 \\
			\end{array}
			\right) \nonumber \\
			&& + \mathrm{i} \omega \left(
			\begin{array}{ccc}
				0 & 0 & 0 \\
				0 & -8 Y Z^2 & 2  Z \left(4 Y^2 - Z\right) \\
				0 & 2 Z \left(4 Y^2 - Z\right) & 4 Y \left( Z - 2 Y^{2} \right)\\
			\end{array}
			\right) \Bigg],\label{gijnn-4}\\
			(\mathcal{G}^{(n,n-3)}_{ij}) &=& \frac{1}{8}\sqrt{\frac{n (n-1) (n-2)}{2 \hbar Z^{3} \omega^{11}}} \Bigg[ \left(
			\begin{array}{ccc}
				0 & 2 Y Z^2 \omega^{2} & Z \omega^{2} \left( Z-2Y^{2} \right) \\
				2 Y Z^2 \omega^{2} & 4 W Z^2 \left(3 Y^2-Z\right) & 2 W Y Z \left(3Z - 5 Y^2\right) \\
				Z \omega^{2} \left( Z-2Y^{2} \right) & 2 W Y Z \left(3Z - 5 Y^2\right) & 2 W Y^{2} \left(4 Y^{2} - 3Z\right) \\
			\end{array}
			\right) \nonumber \\
			&& + \mathrm{i} \omega \left(
			\begin{array}{ccc}
				0 & 2 Z^{2} \omega^{2} & -2 Y Z \omega^{2}\\
				2 Z^{2} \omega^{2} & 12 W Y Z^{2} & -WZ \left(9Z - 10 \omega^{2}\right) \\
				-2 Y Z \omega^{2} & -WZ \left(9Z - 10 \omega^{2}\right) & 2 WY \left(4 Y^{2} - Z\right) \\
			\end{array}
			\right)  \Bigg], \\
			(\mathcal{G}^{(n,n-2)}_{ij}) &=& \frac{\sqrt{n (n-1)}}{2 \hbar Z \omega^{7}}\Bigg[ \left(
			\begin{array}{ccc}
				- Z^{2} \omega^{4} & - 2 W Y Z^{2} \omega^{2} & W Y^{2} Z \omega^{2} \\
				- 2 W Y Z^{2} \omega^{2} & -W^{2} Z^{2} \left(5 Y^{2} - Z\right) & W^{2} Y Z \left(3 Y^{2}-Z\right) \\
				W Y^{2} Z \omega^{2} & W^{2} Y Z \left(3 Y^{2}-Z\right) & W^2 Y^2 \left(Z-2 Y^2\right) \\
			\end{array}
			\right) \nonumber  \\
			&& + \mathrm{i} \omega \left(
			\begin{array}{ccc}
				0 & - W Z^{2} \omega^{2} & WYZ\omega^{2} \\
				- W Z^{2} \omega^{2} & - 4 W^{2} Y Z^{2} & 3 W^{2} Y^{2} Z \\
				WYZ\omega^{2} & 3 W^{2} Y^{2} Z & -2 W^{2} Y^{3} \\
			\end{array}
			\right) \Bigg], \\
			(\mathcal{G}^{(n,n-1)}_{ij}) &=& \frac{n}{8} \sqrt{\frac{n}{2\hbar Z \omega^{9}}} \Bigg[ \frac{1}{Z\omega }  \left(
			\begin{array}{ccc}
				0 & -2 Y Z^2 \omega^{2} & Z \left(2 Y^{2} - Z\right) \omega^{2} \\
				- 2 Y Z^2 \omega^{2} & -4 W Z^2 \left(Y^2+Z\right) & 2 W Y Z \left(Y^2+Z\right) \\
				Z \left(2 Y^{2} -Z\right) \omega^{2} & 2 W Y Z \left(Y^2+Z\right) & -2 W Y^2 Z \\
			\end{array}
			\right)  \nonumber\\
			&& + \mathrm{i} \left(
			\begin{array}{ccc}
				0 & - 2 Z \omega^{2} & 2 Y \omega^{2} \\
				- 2 Z \omega^{2} & -4 W Y Z & W \left(2 Y^{2}+Z\right) \\
				2 Y \omega^{2} & W \left(2 Y^{2}+Z\right) & -2 W Y \\
			\end{array}
			\right) \Bigg],
		\end{eqnarray}
		\begin{eqnarray}
			(\mathcal{G}^{(n,n+1)}_{ij}) &=& \frac{n+1}{8} \sqrt{\frac{n+1}{2\hbar Z \omega^{9}}} \Bigg[ \frac{1}{Z\omega }  \left(
			\begin{array}{ccc}
				0 & -2 Y Z^2 \omega^{2} & Z \left(2 Y^{2} - Z\right) \omega^{2} \\
				- 2 Y Z^2 \omega^{2} & -4 W Z^2 \left(Y^2+Z\right) & 2 W Y Z \left(Y^2+Z\right) \\
				Z \left(2 Y^{2} -Z\right) \omega^{2} & 2 W Y Z \left(Y^2+Z\right) & -2 W Y^2 Z \\
			\end{array}
			\right)  \nonumber\\
			&& - \mathrm{i} \left(
			\begin{array}{ccc}
				0 & - 2 Z \omega^{2} & 2 Y \omega^{2} \\
				- 2 Z \omega^{2} & -4 W Y Z & W \left(2 Y^{2}+Z\right) \\
				2 Y \omega^{2} & W \left(2 Y^{2}+Z\right) & -2 W Y \\
			\end{array}
			\right) \Bigg],\\
			(\mathcal{G}^{(n,n+2)}_{ij}) &=& \frac{\sqrt{(n+1)(n+2)}}{2 \hbar Z \omega^{7}}\Bigg[ \left(
			\begin{array}{ccc}
				- Z^{2} \omega^{4} & - 2 W Y Z^{2} \omega^{2} & W Y^{2} Z \omega^{2} \\
				- 2 W Y Z^{2} \omega^{2} & -W^{2} Z^{2} \left(5 Y^{2} - Z\right) & W^{2} Y Z \left(3 Y^{2}-Z\right) \\
				W Y^{2} Z \omega^{2} & W^{2} Y Z \left(3 Y^{2}-Z\right) & W^2 Y^2 \left(Z-2 Y^2\right) \\
			\end{array}
			\right) \nonumber  \\
			&& - \mathrm{i} \omega \left(
			\begin{array}{ccc}
				0 & - W Z^{2} \omega^{2} & WYZ\omega^{2} \\
				- W Z^{2} \omega^{2} & - 4 W^{2} Y Z^{2} & 3 W^{2} Y^{2} Z \\
				WYZ\omega^{2} & 3 W^{2} Y^{2} Z & -2 W^{2} Y^{3} \\
			\end{array}
			\right) \Bigg],\\
			(\mathcal{G}^{(n,n+3)}_{ij}) &=& \frac{1}{8}\sqrt{\frac{(n+1) (n+2) (n+3)}{2 \hbar Z^{3} \omega^{11}}} \Bigg[ \left(
			\begin{array}{ccc}
				0 & 2 Y Z^2 \omega^{2} & Z \omega^{2} \left( Z-2Y^{2} \right) \\
				2 Y Z^2 \omega^{2} & 4 W Z^2 \left(3 Y^2-Z\right) & 2 W Y Z \left(3Z - 5 Y^2\right) \\
				Z \omega^{2} \left( Z-2Y^{2} \right) & 2 W Y Z \left(3Z - 5 Y^2\right) & 2 W Y^{2} \left(4 Y^{2} - 3Z\right) \\
			\end{array}
			\right) \nonumber \\
			&& - \mathrm{i} \omega \left(
			\begin{array}{ccc}
				0 & 2 Z^{2} \omega^{2} & -2 Y Z \omega^{2}\\
				2 Z^{2} \omega^{2} & 12 W Y Z^{2} & -WZ \left(9Z - 10 \omega^{2}\right) \\
				-2 Y Z \omega^{2} & -WZ \left(9Z - 10 \omega^{2}\right) & 2 WY \left(4 Y^{2} - Z\right) \\
			\end{array}
			\right)  \Bigg],\\
			(\mathcal{G}^{(n,n+4)}_{ij}) &=&  \frac{\sqrt{(n+1) (n+2) (n+3) (n+4)}}{64 Z^{2} \omega^{4}} \Bigg[\left(
			\begin{array}{ccc}
				0 & 0 & 0 \\
				0 & 4 Z^2 \left(Z-2 Y^2\right) & 2 Y Z \left(4 Y^2-3 Z\right) \\
				0 & 2 Y Z \left(4 Y^2-3 Z\right) & -8 Y^4+8 Y^2 Z-Z^2 \\
			\end{array}
			\right) \nonumber \\
			&& - \mathrm{i} \omega \left(
			\begin{array}{ccc}
				0 & 0 & 0 \\
				0 & -8 Y Z^2 & 2  Z \left(4 Y^2 - Z\right) \\
				0 & 2 Z \left(4 Y^2 - Z\right) & 4 Y \left( Z - 2 Y^{2} \right)\\
			\end{array}
			\right) \Bigg] .
	\end{eqnarray}}
	
	Lastly, allow us to work with the torsion, the anti-symmetric components of $M^{(n,m)}_{ij}$. As mentioned before, we can either use~\eqref{Gamma_ejem} and \eqref{T_def} or just \eqref{T_def2} to compute the torsion, both render the same result. Therefore, the only non-zero torsion is for $m=n\pm1$, and they are:
	\begin{eqnarray}
		(T^{(n,n-1)}_{ij}) &=&  \sqrt{\frac{n}{32 \hbar Z \omega^{9}}} \left[
		\left( \begin{array}{ccc}
			0 & -2 Z \omega^{2} & 2 Y \omega^{2}  \\
			2 Z \omega^{2} & 0 & W (Z + 2Y^{2}) \\
			-2 Y \omega^{2} & -W (Z + 2Y^{2}) & 0
		\end{array} \right) \right. \nonumber \\
		&&	\left. + \mathrm{i} \omega \left( \begin{array}{ccc}
			0 & 2 YZ  &  Z - 2Y^{2} \\
			-2YZ  & 0 & 2 WY \\
			-Z + 2Y^{2}  & -2 WY  & 0
		\end{array} \right)	\right] \\
		(T^{(n,n+1)}_{ij}) &=&  \sqrt{\frac{n+1}{32 \hbar Z \omega^{9}}} \left[
		\left( \begin{array}{ccc}
			0 & -2 Z \omega^{2} & 2 Y \omega^{2}  \\
			2 Z \omega^{2} & 0 & W (Z + 2Y^{2}) \\
			-2 Y \omega^{2} & -W (Z + 2Y^{2}) & 0
		\end{array} \right) \right. \nonumber \\
		&& \left.- \mathrm{i} \omega \left( \begin{array}{ccc}
			0 & 2 YZ  &  Z - 2Y^{2} \\
			-2YZ & 0 & 2 WY \\
			-Z + 2Y^{2} & -2 WY & 0 \label{Tnn+1ij}
		\end{array} \right)
		\right].
	\end{eqnarray}
	From the expressions \eqref{gijnn-4} to \eqref{Tnn+1ij}, we observe that the only states that correlate are those that differ by up to four levels and can be connected by moving the system parameters. \\

    Regarding the scalar invariants, in this case, they depend on the parameters that describe the system. However, their behavior is similar to that exposed in the previous example. The invariant $\mathcal{N}^{(n,n \pm 1)}_{T}$ declines as $n$ increases, whereas the rest of the invariants approach non-zero fixed values. The dependency on the parameters also affects the asymptotic value each invariant reaches, except for the torsion invariants, whose value always tends to zero regardless of the parameter choice. \\

	\section{Conclusions}
	\label{Sec_concl}
	
	Employing the geometric principles of Cartan geometry, this study sheds light on the nature of non-adiabatic coupling vectors, offering a deeper understanding of their role in Quantum Mechanics. These vectors are introduced by rewriting $\ket{\partial_{i} n}$ in terms of a complete basis of the system. The part aligned with the $n$-th state corresponds directly to the Berry connection, and the remaining parts correspond to the non-adiabatic coupling vectors, which we call $N$-beins. We opt for this name because it acts as the square root of the quantum geometric tensor, similar to how the vielbein is the square root of the metric in the Cartan formalism.\\
	
	Moreover, our interpretation as $N$-beins allows us to introduce the two-state quantum geometric tensor $M_{ij}^{(n,m)}$, which has a symmetric part $\mathcal{G}_{ij}^{(n,m)}$ similar to a metric; however, it is not real. Additionally, $M_{ij}^{(n,m)}$ has an anti-symmetric part $T_{ij}^{(n,m)}$ that we can interpret as the quantum torsion of the system because it could be written as the covariant derivative of the $N$-bein. In general, the two-state geometric tensor $M_{ij}^{(n,m)}$ measures the amplitude of reaching the state $|m\rangle$ starting from the state $|n\rangle$ by changing the parameters $\lambda^{j}$ and $\lambda^{i}$. We observe that $M^{(n,m)}_{ij}$ is similar in form to the quantum geometric tensor except that we are now working with two distinct $N$-beins that share a common state $\ket{l}$, different from both $|n \rangle$ and $|m \rangle$.\\
 
 Furthermore, the order of variations becomes significant when considering the tensor $M^{(n,m)}_{ij}$. If $T^{(n,m)}_{ij}=0$, then $M^{(n,m)}_{ij}$ is symmetric, and the order does not matter. However, a non-zero $T^{(n,m)}_{ij}$ introduces an anti-symmetric part in $M^{(n,m)}_{ij}$, making the variation sequence (from $|n\rangle$ to $|m\rangle$) crucial. In contrast, the inherently symmetric tensor $\mathcal{G}^{(n,m)}_{ij}$ is independent of the variation order. Thus, $\mathcal{G}_{ij}^{(n,m)}$ determines the path between the states $|n\rangle$ and $|m\rangle$ by requiring the variation of two parameters. In this way, these quantities measure the correlation between two quantum states and how they connect under a variation of the parameters. The examples show how the correlation is selected and that transitions are avoided since there is no possible correlation between certain states. \\
 
    One point to consider is that, unlike the quantum geometric tensor, the introduced quantities $M_{ij}^{(n,m)}$, $\mathcal{G}_{ij}^{(n,m)}$, and $T_{ij}^{(n,m)}$ are not fully gauge-invariant. Nevertheless, using transformations \eqref{tensor_transf}, true gauge invariants can be built as products of these quantities and be directly measurable. We construct these invariants in Sec. \ref{Sec_invs}, and we define two kinds: tensorial invariants, similar to Riemann tensors, and scalar invariants. Scalar invariants compare the magnitude of the distance between two states with the distance obtained by infinitesimally varying the parameters while remaining in the same state. From the examples, we can observe intriguing behaviors of these invariants. The invariant associated with torsion is significant for states near the ground state and approaches zero for excited states, suggesting that it is highly measurable in the transition from the ground state to the first excited state. On the other hand, the invariant associated with $\mathcal{G}^{(n,m)}_{ij}$ maintains essentially the same importance for states near the base or excited states, implying that it is also susceptible to measurement.\\

    Within the experimental context, the application of the $N$-bein has been explored~\cite{Ahn2203}. Thus, we expect the tools developed in this work to be advantageous in the experimental scheme. In particular,  the tensors derived from $M^{(n,m)}_{ij}$ relate two states after variations in the parameters that define the quantum system. Meanwhile, the invariants defined in Sec.~\ref{Sec_invs} are observables that measure the importance of such tensors. Therefore, we hope  our formalism to be valuable for the experimental framework.

\section*{Acknowledgments}
	
	This work was partially supported by DGAPA-PAPIIT Grant No. IN105422. J.R. acknowledges the financial support from Conahcyt under the ``Estancias Posdoctorales por M\'{e}xico 2022 (3)'' program. C. A. V. gratefully acknowledges Conahcyt for his PhD scholarship (662129).
	
	\bibliographystyle{unsrt}
	\section*{References}
	\bibliography{References}

	\appendix
	
	\section{Properties of the geometric objects}
	\label{Ap_prop}
	
	Throughout this paper we have introduced several new geometric objects, so we decided to devote this part to enlist some of the their most important properties. We start with the projectors $\hat{P}^{(n)}:= \hat{\mathds{1}} - \pj{n}{n}$ and $\hat{P}^{(n,m)}:= \hat{\mathds{1}} - \pj{n}{n} - \pj{m}{m}$, which project onto the subspaces orthogonal to $\ket{n}$ and to $\ket{n}$ and $\ket{m}$, respectively. The projectors satisfy the properties
	\begin{eqnarray}
		\hat{P}^{(n)}= \left( \hat{P}^{(n)} \right)^{2}=\left( \hat{P}^{(n)} \right)^{\dagger}, \\
		\hat{P}^{(n,m)}= \hat{P}^{(m,n)} = \left( \hat{P}^{(n,m)} \right)^{2} = \left( \hat{P}^{(n,m)} \right)^{\dagger}.
	\end{eqnarray}
	Under the gauge transformation, both projectors $\hat{P}^{(n)}$ and $\hat{P}^{(n,m)}$ remain invariant. \\ 
	
	Next, we continue with the properties for the $N$-bein $e^{(n)}_{i \; m}$ defined in~\eqref{e_def}. From the normalization condition $\bk{m(\lambda)}{n(\lambda)}=\delta_{mn}$, we derive the identities
	\begin{eqnarray}
		\label{id_n1}
		&\bk{\partial_{i} m}{n} + \bk{m}{\partial_{i} n} = 0, \\
		\label{id_n2}
		&\bk{\partial_{ij} m}{n} + \bk{\partial_{i} m}{\partial_{j} n} + \bk{\partial_{j} m}{\partial_{i} n} + \bk{m}{\partial_{ij} n} = 0.
	\end{eqnarray}
	Thus, from~\eqref{e_def} it is straightforward to prove 
	\begin{equation}
		\label{e_conj}
		\left( e^{(n)}_{i \; m} \right)^{\ast} = e^{(m)}_{i \; n}.
	\end{equation}
	Since the $N$-bein is the projection of the  varied state $\ket{n(\lambda + \delta \lambda)}$ onto $\ket{m(\lambda)}$ due to the variation of parameter $\lambda^{i}$ [see~\eqref{n_proj}], the conjugation property is just a change in the perspective of the state under study. Instead of focus our attention on the state $\ket{n (\lambda + \delta \lambda)}$ projected onto $\ket{m(\lambda)}$, we are studying $\ket{m(\lambda + \delta \lambda)}$ projected onto $\ket{n(\lambda)}$. Hence, if a $\ket{n(\lambda + \delta \lambda)}$ has components in a particular direction $\ket{m(\lambda)}$, i.e., $e^{(n)}_{i\; m} \neq 0$; then $\ket{m(\lambda + \delta \lambda)}$ also has components in the direction $\ket{n(\lambda)}$. Moreover, notice that~\eqref{e_conj} implies that the real and imaginary part of $e^{(n)}_{i \; m}$, respectively $ \theta^{(n)}_{i \; m}$ and $ \eta^{(n)}_{i \; m}$, satisfy
	\begin{eqnarray}
		\theta^{(n)}_{i \; m} = \theta^{(m)}_{i \; n}, \\
		\eta^{(n)}_{i \; m} = -\eta^{(m)}_{i \; n}.
	\end{eqnarray}
	Therefore, the change of perspective is reflected only in the imaginary part of the $N$-bein; the real part is invariant under the change $n \leftrightarrow m$. \\
	
	In Sec.~\ref{Sec_Cartan} we prove that for the gauge transformation $\ket{n}\rightarrow e^{\mathrm{i} \alpha_{n}} \ket{n}$ the $N$-bein transforms similar to the state $\ket{n}$, but with a relative phase $\alpha_{nm}:= \alpha_{n} - \alpha_{m}$, see~\eqref{trule}. It implies that the real and imaginary parts of $e^{(n)}_{i \; m}$ transform as
	\begin{eqnarray}
		\label{trans_theta}
		\left( \theta^{(n)}_{i \; m} \right)^{\prime} =  \theta^{(n)}_{i \; m} \cos \alpha_{nm}  -  \eta^{(n)}_{i \; m} \sin \alpha_{nm}, \\
		\label{trans_eta}
		\left( \eta^{(n)}_{i \; m} \right)^{\prime} = \eta^{(n)}_{i \; m} \cos \alpha_{nm}  +  \theta^{(n)}_{i \; m} \sin \alpha_{nm}.
	\end{eqnarray}
	If the states $\ket{n}$ and $\ket{m}$ transform with the same phase ($\alpha_{n}=\alpha_{m}$), then $\theta^{(n)}_{i \; m}$ and $\eta^{(n)}_{i \; m}$ (and therefore $e^{(n)}_{i \; m}$) are invariant under the gauge transformation. This property is inherit on the subsequent tensors $M^{(n,m)}_{ij}$, $\mathcal{G}^{(n,m)}_{ij}$ and $T^{(n,m)}_{ij}$, see below. On the other hand, we rewrite ~\eqref{trans_theta} and~\eqref{trans_eta} as
	\begin{equation}
		\left( \begin{array}{c} 
			\theta^{(n)}_{i \; m} \\
			\eta^{(n)}_{i \; m} 
		\end{array} \right)^{\prime} = \left( \begin{array}{cc} 
			\cos \alpha_{nm} & -\sin \alpha_{nm} \\
			\sin \alpha_{nm} & \cos \alpha_{nm}
		\end{array} \right) 
		\left( \begin{array}{c} 
			\theta^{(n)}_{i \; m} \\
			\eta^{(n)}_{i \; m} 
		\end{array} \right).
	\end{equation}
	Thus, the gauge transformation is a $SO(2)$ rotation on the real vector $\left( \theta^{(n)}_{i \; m}\quad \eta^{(n)}_{i \; m} \right)^{T}$. Hence, it is no surprise that metric $g^{(n)}_{ij}$ and the curvature $F^{(n)}_{ij}$ are invariant under such transformation since they are related to the norm of the vector $\left( \theta^{(n)}_{i \; m}\quad \eta^{(n)}_{i \; m} \right)^{T}$ and with area enclosed by $\theta^{(n)}_{i \; m}$ and $\eta^{(n)}_{i \; m}$, respectively [see~\eqref{g_forms} and \eqref{F_forms}], being both quantities are invariant under $SO(2)$ rotations.\\
	
	Next, in~\eqref{Gamma_def} we introduced the connection $\Gamma^{(n,m)}_{i}$, it is real and transforms as a connection under the gauge transformation. It also has the property $\Gamma^{(n,m)}_{i} = -\Gamma^{(m,n)}_{i}$, similar to $\eta^{(n)}_{i\; m}$. The connection allows to introduce the torsion $T^{(n,m)}_{ij}$, whose complex conjugation acts similar to the case for $e^{(n)}_{i\; m}$, i.e.,
	\begin{equation}
		\left(T^{(n,m)}_{ij}\right)^{\ast} = T^{(m,n)}_{ij}.
	\end{equation}
	Meanwhile, under the gauge transformation it transforms as
	\begin{equation}
		\label{T_transf}
		\left(T^{(n,m)}_{ij}\right)^{\prime} = e^{\mathrm{i} \alpha_{nm}}T^{(n,m)}_{ij}.
	\end{equation}	
	Thus, it also transforms similar to the $N$-bein, which is to be expected because we constructed $T^{(m,n)}_{ij}$ from the covariant derivative of $e^{(n)}_{i\; m}$.\\
	
	Similarly, for $M^{(n,m)}_{ij}$ and $\mathcal{G}^{(n,m)}_{ij}$ we have the conjugation properties
	\begin{eqnarray}
		\left(M^{(n,m)}_{ij}\right)^{\ast} = M^{(m,n)}_{ji}, \\
		\left(\mathcal{G}^{(n,m)}_{ij}\right)^{\ast} = \mathcal{G}^{(m,n)}_{ij}.
	\end{eqnarray}
	Notice the how the indices change in $M^{(m,n)}_{ij}$, whereas $\mathcal{G}^{(n,m)}_{ij}$ behaves like the torsion under complex conjugation.	In spite of the difference conjugation property, the transformation law for $M^{(n,m)}_{ij}$ is the same as the one for $\mathcal{G}^{(n,m)}_{ij}$ and $T^{(n,m)}_{ij}$, namely
	\begin{eqnarray}
		\label{M_transf}
		\left(M^{(m,n)}_{ij}\right)^{\prime} = e^{\mathrm{i} \alpha_{nm}} M^{(n,m)}_{ij}, \\
		\label{G_transf}
		\left(\mathcal{G}^{(m,n)}_{ij}\right)^{\prime} = e^{\mathrm{i} \alpha_{nm}} \mathcal{G}^{(n,m)}_{ij}.
	\end{eqnarray}
	On the other hand, for the real curvature $R^{(n,m)}_{ij}$ we find $R^{(n,m)}_{ij} = -R^{(m,n)}_{ij}$ and since it is constructed with the curvature $F^{(n,m)}_{ij}$, $R^{(n,m)}_{ij}$ is also invariant under the gauge transformation.\\

\end{document}